\documentclass[aps,pre,twocolumn,superscriptaddress,showpacs,amsmath]{revtex4-1}
\usepackage{graphicx}

\begin{document}

\newcommand{\transp}{\mathsf{T}}

\title{The Average Spectrum Method for Analytic Continuation:\newline Efficient Blocked Modes Sampling and Dependence on Discretization Grid}
\author{Khaldoon Ghanem}
\affiliation{J\"ulich Supercomputer Centre, Forschungszentrum J\"ulich, 52425 J\"ulich, Germany}
\affiliation{Max-Planck-Institut f\"ur Festk\"orperforschung, 70569 Stuttgart, Germany}
\author{Erik Koch} 
\affiliation{J\"ulich Supercomputer Centre, Forschungszentrum J\"ulich, 52425 J\"ulich, Germany}
\affiliation{JARA High-Performance Computing, 52425 J\"ulich, Germany}

\date{\today}
\begin{abstract}
The average spectrum method is a promising approach for the analytic continuation of imaginary time or frequency data to the real axis. It determines the analytic continuation of noisy data from a functional average over all admissible spectral functions, weighted by how well they fit the data. Its main advantage is the apparent lack of adjustable parameters and smoothness constraints, using instead the information on the statistical noise in the data. Its main disadvantage is the enormous computational cost of performing the functional integral. Here we introduce an efficient implementation, based on the singular value decomposition of the integral kernel, eliminating this problem. It allows us to analyze the behavior of the average spectrum method in detail. We find that the discretization of the real-frequency grid, on which the spectral function is represented, biases the results. The distribution of the grid points plays the role of a default model while the number of grid points acts as a regularization parameter. We give a quantitative explanation for this behavior, 
point out the crucial role of the default model 
and provide a practical method for choosing it, 
making the average spectrum method a reliable and efficient technique for analytic continuation.
\end{abstract}

\maketitle

\section{Introduction}
Strongly interacting quantum many-particle problems require non-perturbative solvers. Quantum Monte Carlo (QMC) approaches provide, in the absence of a sign problem, numerically exact results and are therefore widely used. Their key drawback is that they work well 
only for imaginary time or frequency. To make contact with experiment these data have to be analytically continued to obtain the spectral function $A(\omega)$ on the real-frequency axis. This requires solving an integral equation, presenting an ill-posed inverse problem. The standard approach to this problem for strongly correlated electron systems is the Maximum Entropy method (MaxEnt) described in \cite{MaxEntReview96}, which is, with some variations, also used in Eliashberg theory \cite{Tremblay15}  as well as in lattice QCD simulations \cite{QCD13}.

The ill-posedness of the inverse problem implies that the spectral function $A(\omega)$ giving the best fit to the imaginary-axis data in a least-squares sense, while easily determined, is completely useless: It is dominated by rapid oscillations of diverging amplitude, arising from fitting the inevitable statistical noise in the QMC data. The standard approach for overcoming this problem is to impose smoothness on the solution, i.e., to regularize \cite{Hansen}.
The Maximum Entropy method provides a regularization based on Bayesian arguments. It penalizes deviations of the spectral function from a default model, measured by the relative entropy of the two functions. While the non-linearity of the entropy function makes optimization more difficult, it has the important advantage of ensuring the non-negativity of the spectral function. The method provides good results and is so efficient that it is the de facto standard for analytic continuation problems. Still there remains the problem of choosing an appropriate default model and regularization parameter, the latter giving rise to a number of different flavors of MaxEnt \cite{JarrellCorrel12}.

An alternative approach, the Average Spectrum Method (ASM), that promises to avoid these ambiguities was proposed by White \cite{White91} and, independently, in Refs.~\cite{Sandvik98,Gunnarsson07}. The basic idea is of striking elegance: The spectral function is obtained as the average of all physically admissible spectral functions weighted with how well they fit the data given on the imaginary axis. Due to the ill-posedness of the inverse problem there are many spectral functions that differ drastically but fit the data equally well. Taking the average is thus expected to smooth out features that are not supported by the data, providing a regularization without the need for explicit parameters. The practical application of this conceptually appealing approach has, however, so far suffered from the computational cost of its implementations \cite{White91,Sandvik98,Gunnarsson07,Syljuasen08,Fuchs10}.

Here we introduce the blocked modes sampling technique, which overcomes the main limitation of the average spectrum method: The commonly used recipe is to update the sampled spectral function at several points simultaneously, keeping a number of moments of $A(\omega)$ fixed \cite{Sandvik98,Gunnarsson07}. Our more systematic approach introduces global moves, updating not individual components of $A(\omega)$, but changing it at all frequencies at once by an amount proportional to a singular mode of the kernel. This is very efficient when the global moves are not constrained too much by the non-negativity of $A(\omega)$. When the constraint limits these moves significantly it becomes more efficient to partition the frequency axis and perform global moves on the individual frequency blocks. 

Blocked modes sampling makes the average spectrum method fast enough that we can systematically investigate how well it performs the analytic continuation. We find that the results depend on the way the real-frequency axis is discretized: The density function used for picking grid points acts as a default model, i.e., determines the result in the absence of data, while the number of grid points acts as a regularization parameter. That the ASM includes, via the parametrization of the real axis, a default model has already been noticed in \cite{Beach04,Fuchs10}, while in \cite{Sandvik16} it was observed that the results of the ASM are becoming more biased with increasing number of grid points. We find an explanation for this, which provides us with ways to undo the effect of a specific grid. Moreover, we develop a method for judging the reliability of the results of the average spectrum method, making it a reliable approach to analytic continuation.

\section{Average Spectrum Method}

The average spectrum method is designed to solve linear integral equations of the form
\begin{equation}\label{Fredholm}
 g(y) = \int\! K(y,x)\,f(x)\,dx
\end{equation}
for $f(x)$.
Calculating $g(y)$ given $f(x)$ merely involves a numerically stable integration. The inverse problem, on the other hand, is ill-conditioned since it is numerically hard to reconstruct sharp features in $f(x)$ that enter $g(y)$ only after being integrated over. That becomes harder the smoother the kernel $K(y,x)$ as a function of $x$.
The problem is further complicated by the fact that $g(y)$ is usually determined by Monte Carlo methods, i.e., it is only known within the statistical errors of the simulation.

An important application is the determination of the spectral function $A(\omega)$ from the finite-temperature Green function at the fermionic Matsubara frequencies $\omega_m=(2m+1)\pi/\beta$
\begin{equation}
 G(i\omega_m)=\frac{1}{2\pi}\int_{-\infty}^\infty \frac{1}{i\omega_m-\omega}\,A(\omega)\,d\omega \,,
\end{equation}
at imaginary times ($\tau\in(0,\beta)$)
\begin{equation}
 G(\tau) = -\frac{1}{2\pi} \int_{-\infty}^\infty \frac{e^{-\omega\tau}}{1+e^{-\beta\omega}}\,A(\omega)\,d\omega\,,
\end{equation}
or the coefficients $G_l=\sqrt{2l{+}1}\int_0^\beta P_l(2\tau/\beta-1)\,G(\tau)\,d\tau$ of its expansion in Legendre polynomials $P_l(x)$ \cite{Legendre}
\begin{equation}
 G_l = (-1)^{l+1}\frac{\sqrt{2l{+}1}}{2\pi}\frac{\beta}{2}\int_{-\infty}^\infty\frac{i_l^{(1)}(\beta\omega/2)}{\cosh(\beta\omega/2)}A(\omega)\,d\omega
\end{equation}
where $i_l^{(1)}(x)$ are the modified spherical Bessel functions of first kind \cite{DLMF}.

Another important application is the determination of the susceptibility $\chi''(\omega)$ from the correlation function at the bosonic Matsubara frequencies $\omega_m=2m\pi/\beta$
\begin{equation}
 \Pi(i\omega_m) = \frac{2}{\pi}\int_0^\infty \frac{\omega^2}{\omega_m^2+\omega^2}\,\frac{\chi''(\omega)}{\omega}\,d\omega\,,
\end{equation}
imaginary times
\begin{equation}
 \Pi(\tau) = \frac{1}{\pi}\int_0^\infty\omega\,\frac{e^{-\omega\tau}+e^{+\omega\tau}}{1-e^{-\beta\omega}}\,\frac{\chi''(\omega)}{\omega}\,d\omega\,,
\end{equation}
or its Legendre expansion, which vanishes for odd $l$, while for even $l$
\begin{equation}
 \Pi_l = 
 \frac{\sqrt{2l{+}1}}{\pi}\beta\!\int_0^\infty\!\!\!\omega\,\frac{i_l^{(1)}(\beta\omega/2)}{\sinh(\beta\omega/2)}\,\frac{\chi''(\omega)}{\omega}\,d\omega 
 .
 \end{equation}

In all these cases the function $A(\omega)$ or $\chi''(\omega)/\omega$ to be determined is known to be non-negative.

In practice the QMC data is given as a discrete vector $\mathbf{g}=(g_1,\ldots, g_M)^\dagger$ of $M$ data points. 
The mean over $K$ samples is
\begin{equation}
 \overline{\mathbf{g}}=\frac{1}{K}\sum_{k=1}^K \mathbf{g}_k
\end{equation} 
and its statistical uncertainty, when the samples are uncorrelated, is characterized by the covariance matrix 
\begin{equation}
 \mathbf{C}=\frac{1}{K(K-1)}\sum_k(\mathbf{g}_k-\overline{\mathbf{g}})(\mathbf{g}_k-\overline{\mathbf{g}})^\dagger .
\end{equation}
By the central limit theorem, the probability density of measuring $\overline{\mathbf{g}}$ instead of the exact result $\mathbf{g}_\mathrm{exact}$ is proportional to $\exp(-(\overline{\mathbf{g}}-\mathbf{g}_\mathrm{exact})^\dagger\mathbf{C}^{-1}(\overline{\mathbf{g}}-\mathbf{g}_\mathrm{exact})/2)$.

Given some function $f(x)$, it is straightforward to calculate the corresponding $g[f](y)$ by integration, \eqref{Fredholm}, and discretizing it to obtain $\mathbf{g}[f]$. Assuming that $f(x)$ is the exact model, the probability density for measuring $\overline{\mathbf{g}}$ given covariance $\mathbf{C}$ is
\begin{equation}\label{likelihood}
 p(\overline{\mathbf{g}}|f,\mathbf{C}) \propto e^{-\frac{1}{2}(\overline{\mathbf{g}}-\mathbf{g}[f])^\dagger\,\mathbf{C}^{-1}\,(\overline{\mathbf{g}}-\mathbf{g}[f])}
 =: e^{-\frac{1}{2}\chi^2[f]} \,.
\end{equation}

The idea of the average spectrum method is to average all functions $f(x)$ with the probability that they are the exact model, given the measured data $(\overline{\mathbf{g}}, \mathbf{C})$, i.e., to perform the functional integral
\begin{equation}\label{ASM}
 f_\mathrm{ASM}(\overline{\mathbf{g}},\mathbf{C};\,x) = \int\!\mathcal{D}f\,p(f|\overline{\mathbf{g}},\mathbf{C})\;f(x) .
\end{equation}
By Bayes' theorem the posterior probability density is 
\begin{equation}
 p(f|\overline{\mathbf{g}},\mathbf{C}) = \frac{p(\overline{\mathbf{g}}|f,\mathbf{C})\,p(f)}{p(\overline{\mathbf{g}}|\mathbf{C})}\,,
\end{equation}
where the likelihood is given by \eqref{likelihood}, $p(f)$ is the prior probability density, and $p(\overline{\mathbf{g}}|\mathbf{C})=\int\mathcal{D}f\,p(\overline{\mathbf{g}}|f,\mathbf{C})\,p(f)$ is the normalization. 
For the spectral function and susceptibilities we know that $f$ is non-negative. 
Setting the prior probability to zero for models that violate this constraint and constant otherwise, \eqref{ASM} becomes
\begin{equation}\label{PI}
 f_\mathrm{ASM}(\overline{\mathbf{g}},\mathbf{C};\,x) \propto \int_{f(x)\ge0}\!\mathcal{D}f\,e^{-\frac12\chi^2[f]}\;f(x)\,.
\end{equation}
Estimating $f(x)$ just requires performing an integral over non-negative models while there is no need for any adjustable parameters. Instead, the regularization results exclusively from the uncertainty in the data as given by the covariance $\mathbf{C}$: the larger the statistical noise, the stronger the contribution of models that do not fit the data particularly well. We can thus expect that accurate data will give us spectra with sharp features, while for noisy data the spectra will contain less information, being more smoothed out by the averaging \cite{White91,Sandvik98,Gunnarsson07}.

\section{Test cases}\label{TestCases}
For illustrating how the average spectrum method performs we use the test cases introduced in Ref.~\cite{Gunnarsson10}: We try to reconstruct an optical conductivity given by
\begin{equation}\label{model}
 \sigma(\omega) = \frac{1}{1+(\omega/\Gamma_e)^6}\!\! \sum_{p=0,\pm1} \frac{W_{|p|}}{1+((\omega+\mathrm{sgn}(p)\varepsilon_{|p|})/\Gamma_{|p|})^2} 
\end{equation}
where the overall factor with $\Gamma_e=4$ cuts off $\sigma(\omega)$ for large frequencies and the terms in the sum give a (Drude) peak of weight $W_0=0.3$ and width $\Gamma_0=0.3$ (model 1) or 0.6 (model 2), and two symmetric peaks of weight $W_1=0.2$ and width $\Gamma_1=1.2$ centered at $\omega=\pm\varepsilon_1=\pm 3$.
The corresponding correlation function on the bosonic Matsubara frequencies $i\omega_m=2\pi m i/\beta$
\begin{equation}\label{optcond}
 \Pi(i\omega_m) = \frac{2}{\pi}\int_0^\infty d\omega\,\frac{\omega^2}{\omega_m^2+\omega^2}\,\sigma(\omega)
\end{equation}
can be calculated analytically.
The input data for the analytic continuation is the imaginary-frequency correlation function $\Pi_m = \Pi(i\omega_m) (1+r_m)$ on the first 60 Matsubara frequencies $m=0,\ldots,59$ with Gaussian (relative) noise $r_m$ of variance $\sigma_\Pi$, where $\sigma_\Pi=0.01$ (noisy data) or 0.001 (accurate data). The inverse temperature is $\beta=15$.

\section{Blocked Modes Sampling}

To evaluate the functional integral \eqref{PI} numerically, we discretize $f(x)$. Introducing a grid of $N$ intervals, we can, e.g., represent it as a piece-wise constant function of value $f_n$ on interval $n$: $\mathbf{f}=(f_1,\ldots,f_N)^\transp$. The integral equation \eqref{Fredholm} then becomes a linear equation
\begin{equation}\label{matrixeq}
 \mathbf{g} = \mathbf{K}\mathbf{f}
\end{equation}
and the functional $\chi^2[f]$ is approximated by
\begin{equation}\label{chi2}
 \chi^2(\mathbf{f}) = (\overline{\mathbf{g}}-\mathbf{K}\mathbf{f})^\dagger\,\mathbf{C}^{-1}\,(\overline{\mathbf{g}}-\mathbf{K}\mathbf{f}) .
\end{equation}
It is then easy to modify \eqref{chi2} such that the covariance matrix no longer appears explicitly. For this we factorize $\mathbf{C}^{-1}=\mathbf{T}^\dagger\mathbf{T}$, e.g., by Cholesky decomposition, to obtain 
\begin{equation}
 \chi^2(\mathbf{f}) = (\tilde{\mathbf{g}}-\tilde{\mathbf{K}}\mathbf{f})^\dagger\,(\tilde{\mathbf{g}}-\tilde{\mathbf{K}}\mathbf{f})
 = \|\tilde{\mathbf{g}}-\tilde{\mathbf{K}}\mathbf{f}\|^2
\end{equation}
with $\tilde{\mathbf{g}}:=\mathbf{T}\overline{\mathbf{g}}$ and $\tilde{\mathbf{K}}=\mathbf{TK}$. The covariance $\tilde{\mathbf{C}}$ of the transformed data $\tilde{\mathbf{g}}$ is, by construction, the unit matrix.

The functional integral \eqref{PI} is then estimated from
\begin{equation}\label{MCint}
 \mathbf{f}_\mathrm{ASM}(\tilde{\mathbf{g}}) \propto \prod_{n=1}^N \int_0^\infty\!\! df_n\,\mathbf{f}\,e^{-\frac12\chi^2(\mathbf{f})}\,.
\end{equation}
This $N$-dimensional integral can be evaluated by Monte Carlo techniques.

\subsection{Components Sampling}
The straightforward method for evaluating \eqref{MCint} is to perform a random walk in the space of non-negative vectors $\mathbf{f}$, updating a single component, $f_n\to f_n'$, at a time. Detailed balance is fulfilled if we sample $f_n'$ from the conditional distribution $\propto\exp(-\chi^2(\mathbf{f}; f_n')/2)$ with
\begin{align}
 \chi^2(\mathbf{f}; f_n') &= \big\|\underbrace{\tilde{\mathbf{g}}-\tilde{\mathbf{K}}\mathbf{f}}_{=:\tilde{\mathbf{r}}} - \tilde{\mathbf{K}}_n (f_n'{-}f_n)\big\|^2\label{CompChi}\\
 \nonumber
 &= \tilde{\mathbf{K}}_n^\dagger\tilde{\mathbf{K}}_n\!\left(\!f_n'-f_n{-}\frac{\Re\tilde{\mathbf{K}}_n^\dagger\tilde{\mathbf{r}}}{\tilde{\mathbf{K}}_n^\dagger\tilde{\mathbf{K}}_n}\right)^2 
 \!\!+ \tilde{\mathbf{r}}^\dagger\tilde{\mathbf{r}}{-}\frac{(\Re\tilde{\mathbf{K}}^\dagger_n\tilde{\mathbf{r}})^2}{\tilde{\mathbf{K}}^\dagger_n\tilde{\mathbf{K}}_n}
\end{align}
where $\tilde{\mathbf{K}}_n$ is the $n$-th column of $\tilde{\mathbf{K}}$.
We thus have to sample $f_n'$ from a univariate Gaussian of width $\sigma=1/\|\tilde{\mathbf{K}}_n\|$ centered at 
$\mu=f_n+\Re\tilde{\mathbf{K}}^\dagger_n\tilde{\mathbf{r}}/\|\tilde{\mathbf{K}}_n\|^2$ and truncated to the non-negative values $f_n'\in[0,\infty)$. This can be done very efficiently \cite{TruncNormal95}.

\begin{figure}
 \center
 \includegraphics[width=0.95\columnwidth]{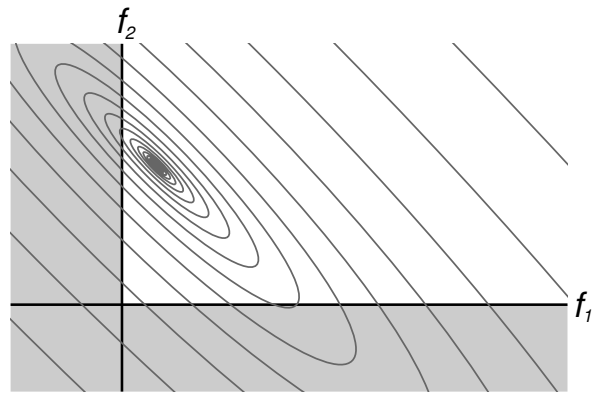}
 \caption{\label{gaussian} Schematic contour plot of the Gaussian probability-density $\exp(-\chi^2(\mathbf{f})/2)$ in the plane of two values $f_1$ and $f_2$. The unphysical region $\mathbf{f}<0$ is shaded in gray. In components sampling the moves $f_i\to f_i'$ are proposed parallel to the coordinate axes, resulting in narrow Gaussians of widths that are of the order of $1/\mathrm{max(d_i)}$. In modes sampling, moves $e_i\to e_i'$ are proposed along the principal axes of the multivariate Gaussian, so that the moves in directions corresponding to small singular values can take large steps. Note that for ill-conditioned problems the singular values $d_i$ vary over many orders of magnitude.}
\end{figure}

Still, sampling components can be very slow because the width of the Gaussian is, in general, extremely small, i.e., the random walk performs only exceedingly small steps. This is evident when sampling spectral functions: we cannot change just a single $f_n$ without violating the sum-rule. A common way out is to update several components simultaneously under the constraint that, e.g., a number of moments of $\mathbf{f}$ is conserved, and to use tempering techniques \cite{White91,Sandvik98,Gunnarsson07,Syljuasen08,Fuchs10}. A simpler and more systematic way is to sample along the principal axes of the multivariate Gaussian $\exp(-\chi^2(\mathbf{f})/2)$, i.e., to change basis. This is illustrated in Fig.~\ref{gaussian}.

\subsection{Modes Sampling}
To implement moves along the principal axes of $\chi^2$, we use the singular value decomposition of the kernel $\tilde{\mathbf{K}}=\mathbf{UDV}^\transp$, where $\mathbf{U}$ is a unitary matrix whose column vectors, $\mathbf{U}_m$, define a basis in the $M$-dimensional data space, $\mathbf{V}$ is a unitary matrix whose columns, $\mathbf{V}_n$, define a basis in the $N$-dimensional space of discretized models, and $\mathbf{D}$ is an $M\times N$ diagonal matrix with diagonal elements $d_1\ge d_2\ge\cdots\ge d_{\mathrm{min}(N,M)}\ge0$. The singular values $d_n>0$ determine how a mode in model space affects the data: $\tilde{\mathbf{K}}\mathbf{V}_n = d_n\mathbf{U}_n$, while the zero modes $\mathbf{U}_n$ with $d_n=0$ or $n>M$  do not affect the data. To simplify the notation we define $d_n:=0$ for $n=\mathrm{min}(N,M){+}1,\ldots,\mathrm{max}(N,M)$.

Transforming to the new bases $\mathbf{h}{:=}\mathbf{U}^\dagger\tilde{\mathbf{g}}$ and $\mathbf{e}{:=}\mathbf{V}^\transp\mathbf{f}$, diagonalizes the quadratic form
\begin{equation}
 \chi^2(\mathbf{f}) = \big\|\mathbf{U}^\dagger\tilde{\mathbf{g}}-\mathbf{DV}^\transp\mathbf{f}\big\|^2
 = \sum_{i=1}^M \big(h_i-d_i\,e_i\big)^2
\end{equation}
and we can write \eqref{MCint} as $\mathbf{f}_\mathrm{ASM}(\tilde{\mathbf{g}})=\mathbf{Ve}_\mathrm{ASM}(\mathbf{h})$, where the integral in the new basis factorizes
\begin{equation}
 \mathbf{e}_\mathrm{ASM}(\mathbf{h})_i \propto \int_{\mathbf{f}\ge0} de_i\,e_i\,
 \exp\big({-}(d_i e_i-h_i)^2/2\big) .
\end{equation}
For evaluating the integral we perform a random walk, now updating one mode $e_i\to e_i'$ at a time. When the corresponding singular value does not vanish, we sample $e_i'$ from a univariate Gaussian of width $\sigma=1/d_i$ centered at $h_i/d_i$ while for $d_i=0$ we sample from a flat distribution. In both cases the distribution is truncated to the interval for which $\mathbf{f}'\ge0$.

Without the non-negativity constraint, the components of $\mathbf{e}_\mathrm{ASM}(\mathbf{h})$ for $d_i>0$ would be given by     $h_i/d_i$, resulting in a least-squares solution that, in general, would be completely dominated by the noise in data modes $h_i$ with exceedingly small singular values. The coupling of the modes through the global condition $\mathbf{f}\ge0$ is thus crucial for regularization.

We find the allowed values of $e_i'$ from the condition $\mathbf{f}'=\mathbf{f}+(e_i'-e_i)\mathbf{V}_i\ge0$, which, in terms of the components, is equivalent to $e_i' \ge e_i - f_n/V_{ni}$ for $V_{ni}>0$ and correspondingly for $V_{ni}<0$. Thus $e_i'$ is constrained by 
\begin{equation}\label{modeLimits}
 \mathrm{max}_n\left\{\left.\frac{f_n}{V_{ni}}\right|V_{ni}<0\right\} 
 \le e_i-e_i' \le
 \mathrm{min}_n\left\{\left.\frac{f_n}{V_{ni}}\right|V_{ni}>0\right\}.
\end{equation}

Sampling modes $e_i'$ is usually much more efficient than sampling components $f_n'$: For modes with large singular value, the Gaussian is narrow so that the random walk quickly jumps close to the expected value $h_i/d_i$ corresponding to the best fit, and then stays close to it. For modes with small or zero singular value the distribution is very broad so that the random walk can take large steps, allowing for an efficient sampling of the degrees of freedom that are not strongly supported by the data. 

Still, sampling may become quite inefficient when non-negativity restricts $e_i'$ to a narrow interval. This will happen when $\mathbf{f}$ has regions where the $f_n$ are very small. 
For a mode $\mathbf{V}_i$ that changes sign on such a region, $e_i'$ cannot differ much from $e_i$ without violating \eqref{modeLimits}. Since the modes form a basis, there are many such modes. In particular, modes sampling can become quite slow when sampling spectral functions on grids with large cutoff. In the tail of the spectral function, where there are many small values $f_n$, it can be more efficient to sample the components $f_n$ directly since they tend to change $\chi^2$, Eq.~\eqref{CompChi}, only little.

\subsection{Blocked Modes Sampling}

The reason for the slow-down of modes sampling is that the narrow intervals originating from regions where the $f_n$ are small also limit the changes in regions where they are large, i.e., where large steps could be taken. We can avoid this by decoupling such regions and sampling them separately. To do this, we split the kernel matrix $\mathbf{K}$ into blocks corresponding to the different regions, perform an SVD for each of them, and sample the resulting blocked modes. Now the non-negativity constraint \eqref{modeLimits} involves only components in the same region. Thus the intervals over which the blocked modes can be sampled will be larger than in modes sampling. On the other hand, the blocked modes no longer give the principal axes of the fit function $\chi^2$ so that the Gaussians from which the modes are sampled will be more narrow than in modes sampling.
When we choose the regions as just the individual grid points we are back to components sampling, where the intervals are semi-infinite $f_n\in[0,\infty)$, while the Gaussians become quite narrow.

\begin{figure}
\includegraphics[width=\columnwidth]{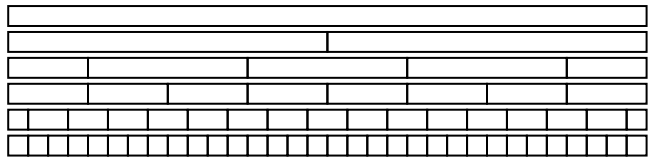}
\caption{\label{blocks} Example of the hierarchy of grid partitionings used in blocked modes sampling. At the highest level (top) the grid on which $\mathbf{f}$ is represented forms a single block. Sampling on this block is modes sampling. 
At the level below the grid is split into two blocks. If going to a lower level we split the blocks in half, there would always be a block boundary at the center of the grid. To avoid this, we shift the intervals at every other level by half their width. At the lowest level (bottom) the blocks are the individual intervals $f_n$. Sampling on these blocks is components sampling.}
\end{figure}

The idea of blocked modes sampling is thus to exploit this trade-off between wide Gaussians and large intervals by interpolating between the limits of modes and components sampling. In practice we use a hierarchy of partitionings of the grid as shown in Fig.~\ref{blocks} and sample in each step all blocks of a randomly chosen hierarchy level.

\subsection{Efficiency}
The computational complexity of the sampling methods per Monte Carlo step are comparable.
For components sampling, calculating the Gaussian parameters, Eq.~\eqref{CompChi}, for updating $f_n\to f_n'$ scales as $\mathcal{O}(MN)$ and there are $N$ components to be updated. 
In modes sampling, the Gaussian parameters are given by the singular values, which are calculated only once, at the beginning of the simulation. Determining the constraint intervals, Eq.~\eqref{modeLimits}, takes $\mathcal{O}(N)$ operations, and there are $N$ modes to be updated. In blocked modes sampling the singular value decompositions for all blocks are calculated once at the beginning. The computational cost of this is dominated by the SVD for the full block and scales as $\mathcal{O}(M\,N^2)$ when there are more grid than data points, $N>M$. 
Sampling a block of length $N/B$ takes $\mathcal{O}(N/B)$ operations for determining the constraint intervals on the block plus $\mathcal{O}(MN(1-1/B))$ operations to calculate the contribution of the other blocks to the Gaussian parameters for each of the $N/B$ modes in the block. There are $B$ such blocks to be updated.
Thus, the computational cost per Monte Carlo step is similar for all three approaches, 
so that their efficiency depends on how much the model $\mathbf{f}$ is changed per MC step.

\begin{figure}
 \includegraphics[width=0.94\columnwidth]{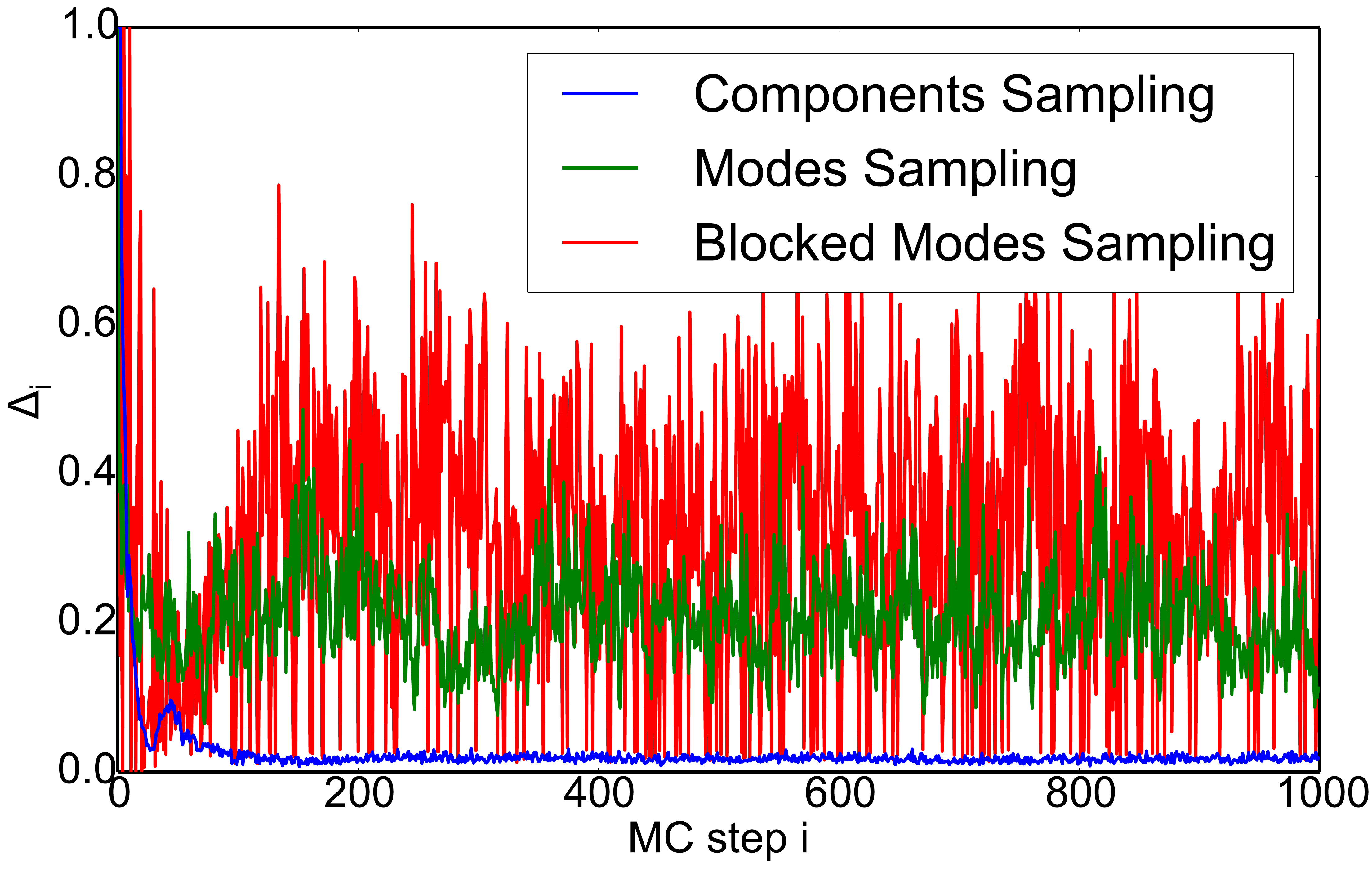}
 \includegraphics[width=0.94\columnwidth]{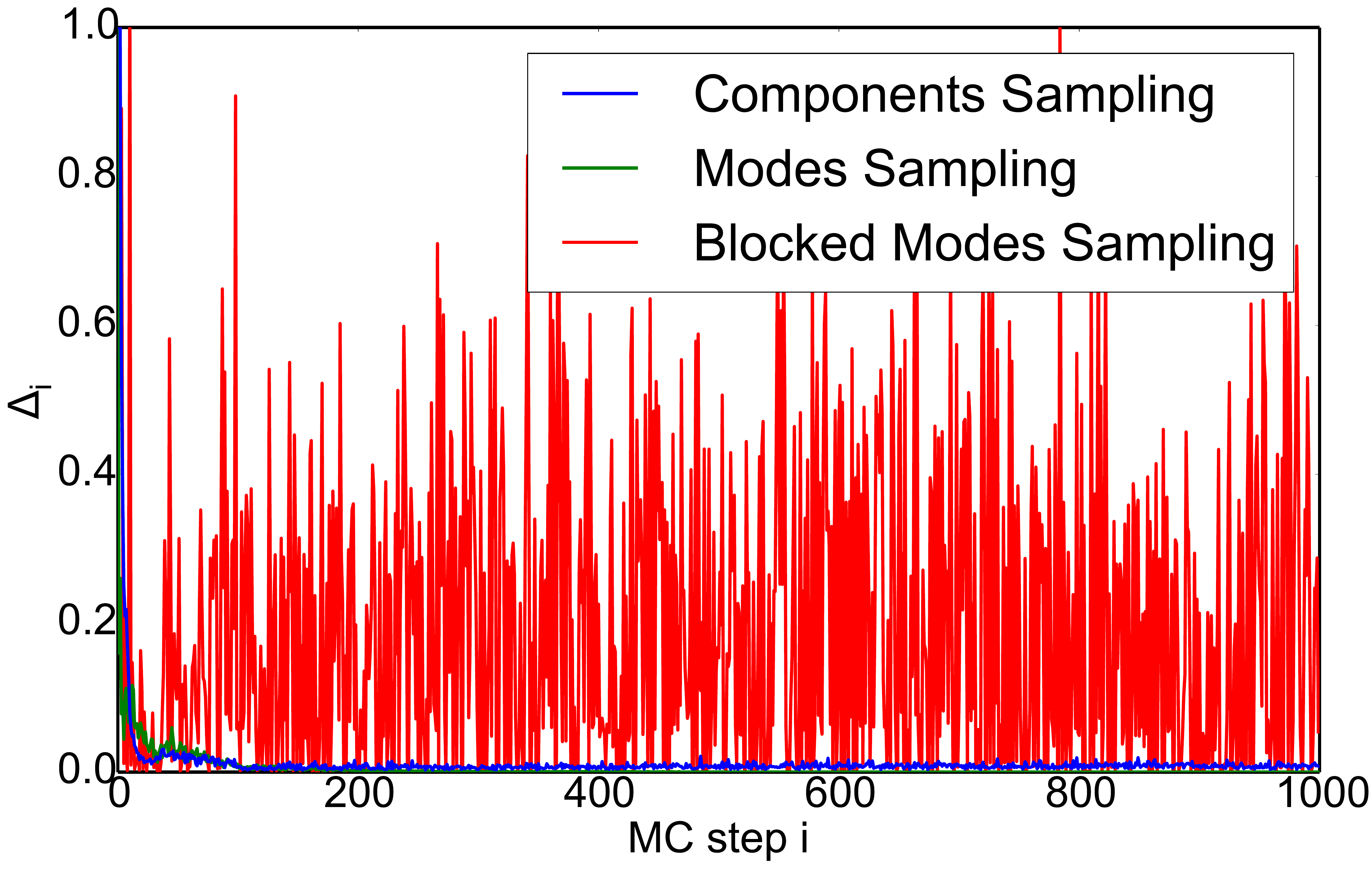}\\[-2ex]
 \caption{\label{MCsteps}(Color online) Efficiency of different sampling methods comparing the changes in $\mathbf{f}$ between Monte Carlo steps, defined as $\Delta_i=\|\mathbf{f}^{(i+1)}-\mathbf{f}^{(i)}\|/\|\mathbf{f}^{(i)}\|$. For a grid with small cutoff (top panel), modes and blocked modes sampling are of comparable efficiency, while the steps taken when sampling components are exceedingly small. When the cutoff is large enough so that the tail of the spectral function is represented on the grid (bottom panel) the steps taken when sampling modes also become extremely small. Only blocked modes sampling always efficiently samples the space of models $\mathbf{f}$.}
\end{figure}
For a practical comparison of the different approaches, we apply them to the test cases, Sec.~\ref{TestCases}: We try to recover the optical conductivity \eqref{model} from \eqref{optcond} for accurate data ($\sigma=0.001$) generated from model 1 ($\Gamma_0=0.3$) on an equidistant frequency grid $\omega_n=0.1(n{-}1/2)$ with small ($n=1,\ldots,40$) and large ($n=1,\ldots,120$) cutoff. The first cutoff is so small (about the width of the overall factor $\Gamma_e$) that the tail of the optical conductivity is hardly represented on the grid. The cutoff for the second grid is chosen such that it covers a large region of the tail where the model is going to zero. 

In both cases, blocked modes sampling updates $\mathbf{f}$ most efficiently so that we obtain uncorrelated samples after only a few Monte Carlo steps. In components sampling $\mathbf{f}$ is hardly changed in a MC update, so that very many steps are needed to obtain statistically independent samples. Modes sampling is as efficient as blocked modes sampling when the model $\mathbf{f}$ does not go to zero. In case the tail is represented on the grid, however, it can become even less efficient than components sampling. 
Fig.~\ref{MCsteps} shows that not every MC step in blocked modes sampling results in a large change in $\mathbf{f}$. Since the level in the hierarchy of blockings (Fig.~\ref{blocks}) is chosen randomly, there are steps where components or the modes of the full grid are updated. But most of the time a blocking in between these extremes is chosen, leading, on average, to an extremely rapid random walk in the space of models. 

Since blocked modes sampling moves so efficiently, it is not very important from which initial vector $\mathbf{f}^{(0)}$ the simulation is started. Still, a good choice is to  start from non-negative least-squares (NNLS) solution \cite{LawsonHanson} of Eq.~\eqref{matrixeq}, since this is gives the best fit under the constraint $\mathbf{f}\ge0$. An even better starting point is obtained by choosing the NNLS solution of Eq.~\eqref{matrixeq}, after adding some noise 
to the data. This moves the initial vector $\mathbf{f}^{(0)}$ slightly away from the best fit solution, such that in effect we can immediately take data without having to warm-up the Monte Carlo run.

\subsection{Linear Constraints}
Besides being non-negative, spectral functions can fulfill other constraints, e.g., the sum rule $\int\!d\omega A(\omega)=2\pi$. 
After discretization such linear constraints can be written as $\mathbf{C\,f}=\mathbf{c}$. For $C$ independent constraints, $\mathbf{C}$ is a $C\times N$ matrix. Using the reduced singular value decomposition $\mathbf{C}=\mathbf{U}_C\mathbf{D}_C\mathbf{V}_C^\transp$, we see that the constraint is only active in the $C$-dimensional subspace that $\mathbf{P}_C=\mathbf{V}_C\mathbf{V}_C^\transp$ projects to.
Fixing $\mathbf{P}_C\mathbf{f}$ to fulfill the constraints, we can sample in the orthogonal space using the methods discussed above.
In practice, we find that sum-rules are strongly represented in the data so that it is not really necessary to enforce them explicitly.

\section{Role of the grid} 
To implement the functional integral \eqref{PI} numerically, we discretize the models $f(x)$ as a finite vector $\mathbf{f}$ representing $f(x)$ on a grid. We now analyze how the results depend on this discretization. 
As test cases we use again the optical conductivity described in Sec.~\ref{TestCases}.

\subsection{Uniform grid}
The most natural choice is to represent $\sigma(\omega)$ on a uniform grid $\omega_n = \omega_0+n\,\Delta\omega$. The  number of grid points $n\in\{1,\ldots,N\}$ must be finite, so that such a grid necessarily has a cutoff. Since the optical conductivity quickly goes to zero for large frequencies, we would expect that once the cutoff is large enough so the tail of $\sigma(\omega)$ is well represented, the result should hardly change when increasing the cutoff further while keeping the step width $\Delta\omega$ fixed.

\begin{figure}
 \includegraphics[width=0.99\columnwidth]{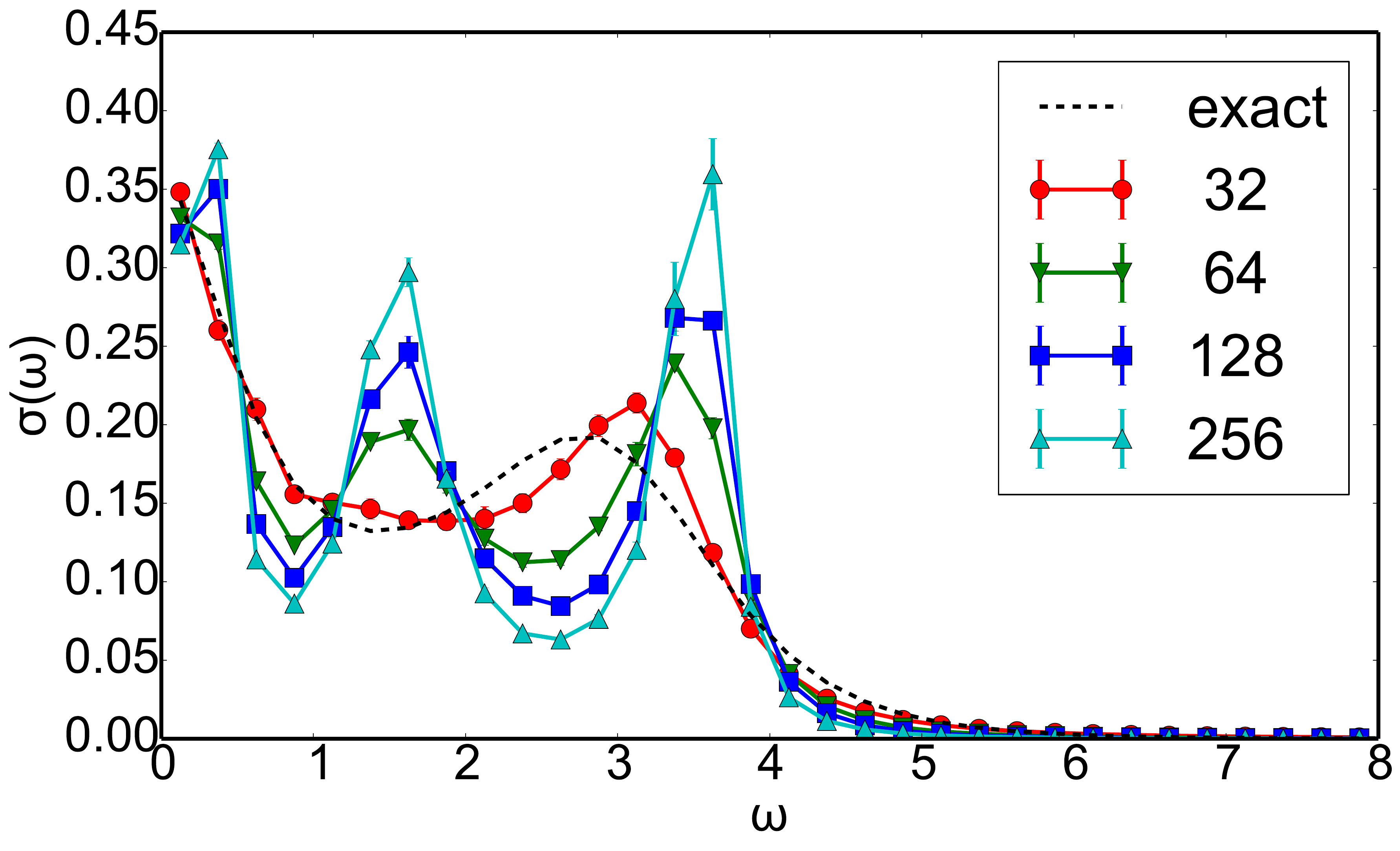}\\[-2ex]
 \caption{\label{cutoff_dependence}(Color online) Optical conductivity $\sigma(\omega)$ obtained by analytic continuation to uniform grids $\omega_n=(n{-}1/2)\Delta\omega$ for $n\in\{1,\ldots,N\}$ with fixed grid spacing $\Delta\omega=0.25$ and increasing number of grid points $N$=~32, 64, 128, and 256, corresponding to a cutoff $\omega_\text{max}\approx$~8, 16, 32, and 64. For comparison, the dashed line shows the optical conductivity from which the imaginary-frequency data for the analytic continuation was calculated. Even though all functions are essentially zero for $\omega\,{\scriptstyle\gtrsim}\,8$, the result depends very strongly on the length of the grid: the result of the analytic continuation develops spurious peaks that get sharper with increasing cutoff.} 
\end{figure}

With our efficient blocked modes sampling we can easily check this. 
For the optical conductivity test cases of Sec.~\ref{TestCases} on grids with $\Delta\omega=0.25$ and $N$=~32, 64, 128, and 256 frequency points, it is a matter of seconds on a modern laptop to obtain the average spectra with good statistical accuracy. The result for model 2 with noise $\sigma_\Pi=$~0.001 is shown in Fig.~\ref{cutoff_dependence}. To our great surprise, we find that the results change drastically: with increasing cutoff a set of pronounced spurious peaks develops.
For the more noisy data, $\sigma_\Pi=$~0.01, the effect gets even stronger.

\subsection{Non-uniform grids}
\begin{figure}
 \includegraphics[width=0.95\columnwidth]{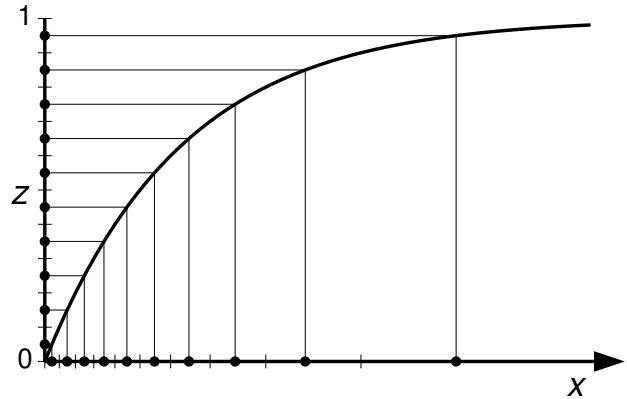}\\[-0.5ex]
 \caption{\label{grid_mapping}Grid mapping from uniform grid on the interval $[0,1]$ to a non-uniform grid on the half infinite interval $[0,\infty)$. Grid points are indicated as dots, limits of intervals by bars.}
\end{figure}

To eliminate the cutoff for a finite grid on an infinite interval we need to choose the grid points such that their spacing increases with their value. We can construct such a grid $x_n$ on a general interval $x_\mathrm{min}\ldots x_\mathrm{max}$ using a positive and normalized function $\rho(x)$ that defines the density of the grid points. The cumulative distribution function $P(x):=\int_{x_\mathrm{min}}^x\!dx'\rho(x')$ is then a monotonous function 
mapping the interval $x_\mathrm{min}\ldots x_\mathrm{max}$ to $[0,1]$. Choosing a uniform discretization $z_n=(n{-}1/2)/N\in[0,1]$ with $n\in\{1,\ldots,N\}$ we obtain a grid $x_n=P^{-1}(z_n)$. 
To get a more intuitive notation, we write the cumulative distribution function as $z(x):=P(x)$ and its inverse as $x(z):=P^{-1}(z)$. Then the $x$-grid $x_n=x(z_n)$ is given in terms of the uniform $z$-grid. This mapping is illustrated in Fig.~\ref{grid_mapping} for the interval $0\ldots\infty$.

The following table lists a few useful non-uniform grids for the semi-infinite interval $[0,\infty)$. The names for the grids are derived from their density function. Note that our exponential grid is also known as logarithmic mesh, while our Lorentzian grid is sometimes called a conformal parametrization \cite{Rubtsov06,Millis16}. 
For the Gaussian grid, $\mathrm{inverf}$ is the inverse of the error function \cite{DLMF}.
\begin{equation*}
\begin{array}{lccc}
& 
\displaystyle\rho(x)=\frac{dz}{dx} & 
x(z) & 
\displaystyle\frac{dx}{dz}\\[2ex]\hline\\[-2ex]
\mbox{Gaussian} & 
\displaystyle\frac{e^{-x^2/2\alpha^2}}{\sqrt{2\pi}\alpha/2} & 
\sqrt{2}\alpha\,\mathrm{inverf}(z) &
\displaystyle\frac{\sqrt{2\pi}\alpha/2}{e^{-\mathrm{inverf}(z)^2}}
\\[2.5ex]
\mbox{exponential} & 
\displaystyle\frac{e^{-x/\beta}}{\beta} & 
-\beta\ln(1{-}z) & 
\displaystyle\frac{\beta}{1-z}
\\[2.5ex]
\mbox{Lorentzian} &
\displaystyle\frac{2/\pi\gamma}{1+(x/\gamma)^2} &
\gamma\tan(\pi z/2) &
\displaystyle\frac{\pi\gamma/2}{\cos(\pi z/2)^2}
\\[1ex]
\end{array}
\end{equation*}

The Gaussian and Lorentzian grids are easily extended to the interval $(-\infty,\infty)$ by replacing $z$ by $2z-1$, giving $x_\text{Gau\ss}(z)=\sqrt{2}\alpha\,\mathrm{inverf}(2z-1)$ and $x_\mathrm{Lor}(z)=-\gamma\cot(\pi z)$.

We express the integral equation 
in the new variable
\begin{equation*}
 g(y)=\!\int\!K(y,x)\,f(x)\,dx = \!\int_0^1\!\!K(y,x(z))\,f(x(z))\frac{dx}{dz}\,dz .
\end{equation*}
To obtain a matrix equation as in \eqref{matrixeq} we write the integral as a Riemann sum \cite{Waldvogel}
\begin{equation}\label{Riemann}
 g(y)\approx\frac{1}{N}\sum_{n=1}^{N} K(y,x_n)\,f(x_n)\,\frac{dx(z_n)}{dz}\,.
\end{equation}
Since $w_n:=(1/N)\,dx(z_n)/dz=1/N\rho(x(z_n))$ is approximately the width of the interval $[x(z_{n-1/2}),\,x(z_{n+1/2})]$, 
we can interpret $\bar{f}_n:=f(x_n)\,w_n$ as the integral of $f(x)$ over that interval.

\begin{figure}
 \includegraphics[width=0.94\columnwidth]{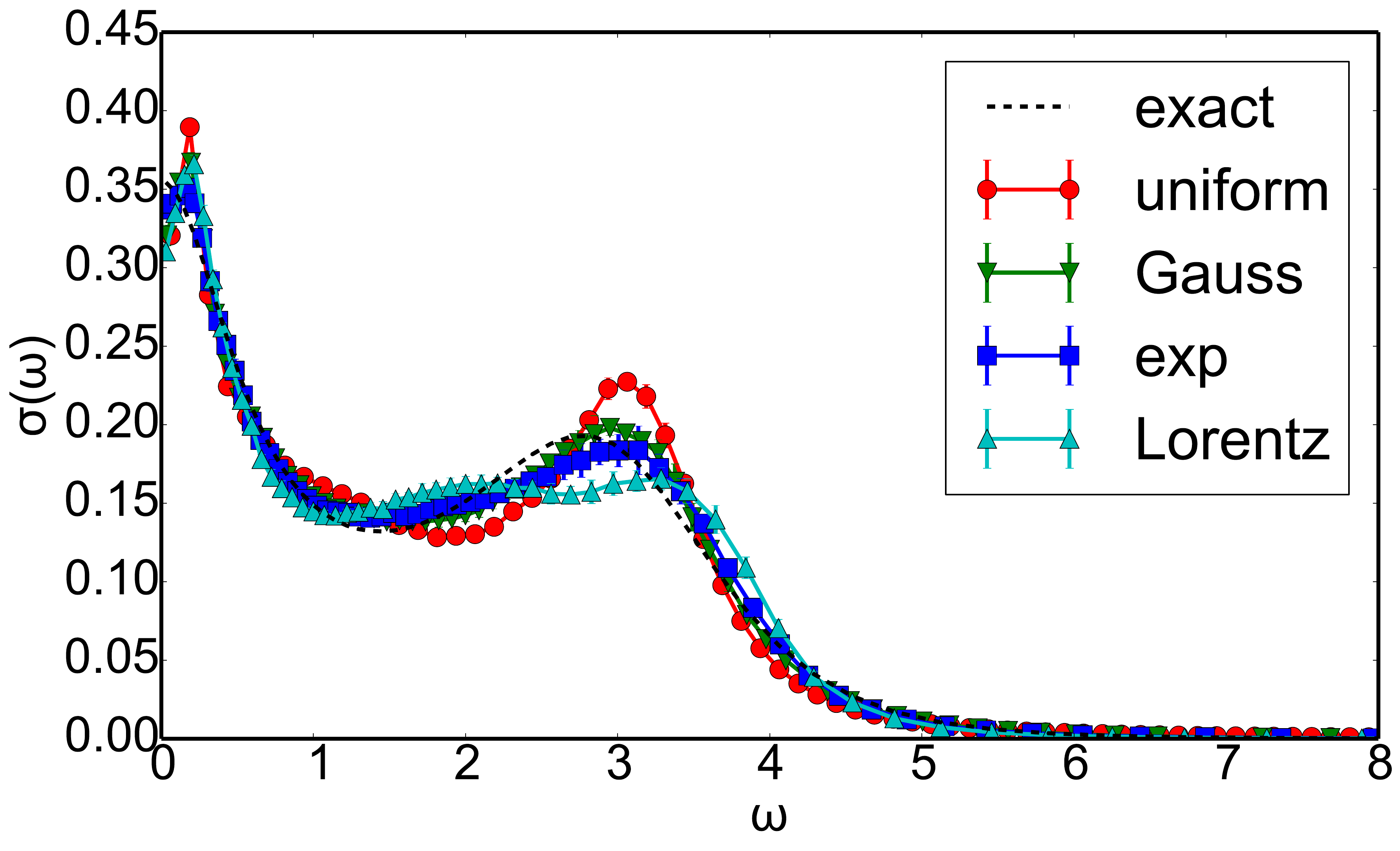} 
 \includegraphics[width=0.94\columnwidth]{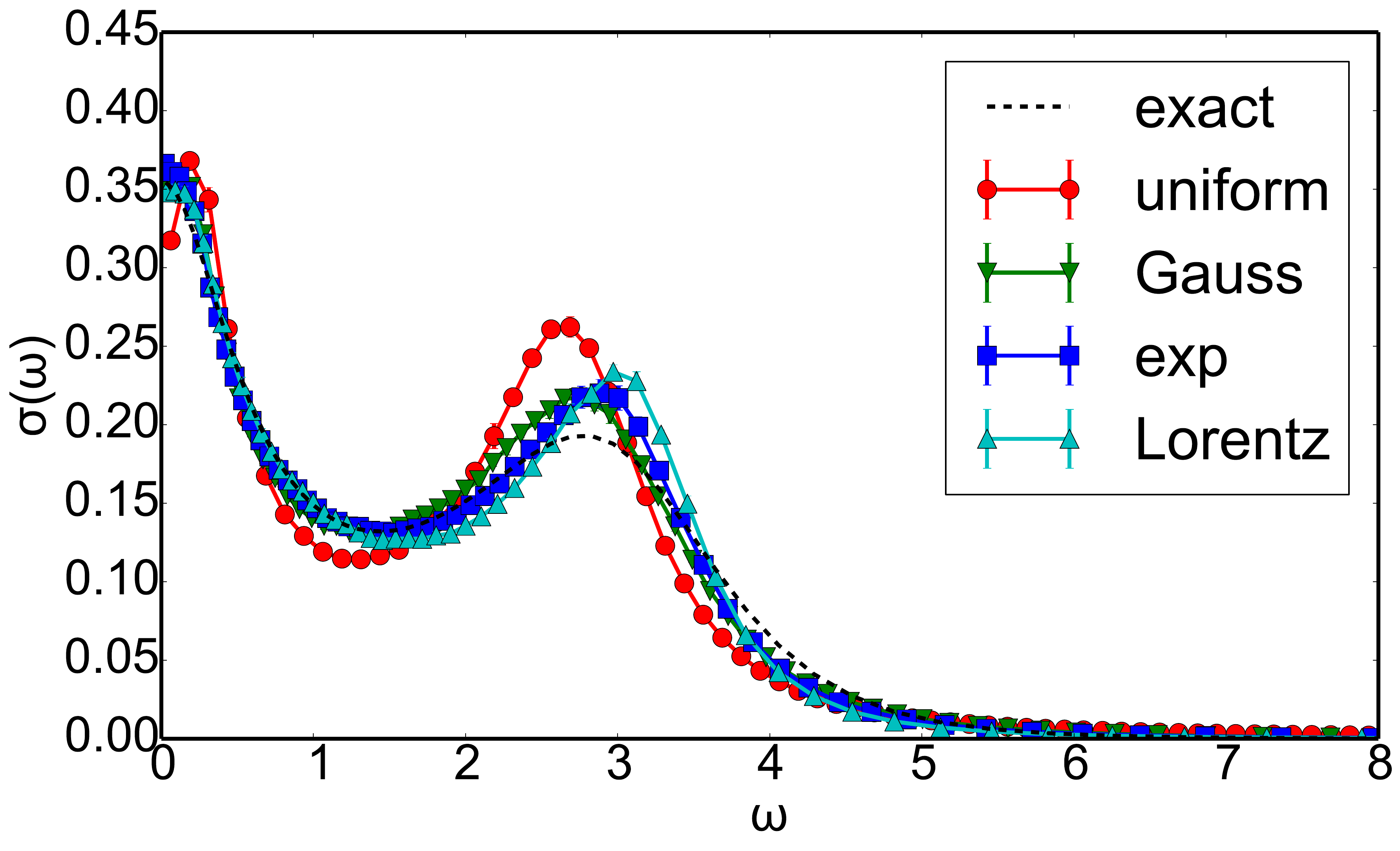}\\[-2ex]  \caption{\label{grid_dependence}(Color online) Optical conductivity $\sigma(\omega)$ obtained by analytic continuation to different grids of $N=64$ points. The uniform grid has a spacing $\Delta\omega=0.125$, corresponding to a cutoff $\omega_\mathrm{max}\approx8$. The width parameter of the Gaussian grid was chosen $\alpha=4$, for the exponential $\beta=3$, and for the Lorentzian $\gamma=2.5$. The dashed line shows the exact result. Removing the cutoff by going to a non-uniform grid improves the result significantly, and the results depend much less on the chosen width parameter than on the cutoff. Still, the average spectra obtained for different grid densities differ by more than their error bars. This grid dependence becomes somewhat stronger for larger noise in the data (upper panel $\sigma_\Pi=0.001$, lower panel $\sigma_\Pi=0.01$).}
\end{figure}
Writing the matrix form of the integral equation as $\mathbf{g}=\mathbf{K}\bar{\mathbf{f}}$, we perform the integral (cf.~\eqref{MCint}) over the $\bar{\mathbf{f}}$. The results for model 2 of Sec.~\ref{TestCases} are shown in Fig.~\ref{grid_dependence}. We find that using non-uniform grids tends to give a dramatic improvement over the results for uniform grids with cutoff (Fig.~\ref{cutoff_dependence}). Still, results do depend on the choice of the grid, the more so the larger the noise in the data.

We can understand this by considering the limit where the data contains no information about the model except a sum rule $\sum\bar{f}_n=1$ to keep the result finite. Then \eqref{MCint} becomes
\begin{equation}\label{def}
 \bar{\mathbf{f}}_\mathrm{ASM}\propto\prod_{n=1}^N\int_0^\infty\!d\bar{f}_n\,\bar{\mathbf{f}}\;\,\delta\!\left(\sum_{n=1}^N\bar{f}_n-1\right) \!.
\end{equation}
In this integral, all $\bar{f}_n$ play the same role, so that by symmetry all components of $\bar{\mathbf{f}}_\mathrm{ASM}$ must be the same and, by the sum rule, equal to $1/N$. Consequently, in the absence of data except for a sum-rule, the average spectrum is equal to $\mathbf{f}_\mathrm{ASM}(x_n) = 1/N w_n =\rho(x_n)$. In that sense, the grid density acts as a default model.

In the average spectra of Fig.~\ref{grid_dependence} the effect of the grid is most clearly seen in the way the tail goes to zero. The Fredholm integral for the optical conductivity \eqref{optcond}, e.g., depends, except for the sum rule given by $\Pi(0)$, only very weakly on the form of $\sigma(\omega)$ at large frequencies,
\begin{equation}
    \Pi(i\omega_m)-\Pi(0) = -\frac{2}{\pi}\int_0^\infty\!d\omega\,\frac{\sigma(\omega)}{1+(\omega/\omega_m)^2} \,,
\end{equation}
so that the data contains only little information about the shape of the tail. Indeed, as expected from \eqref{def}, we find that for large 
$\omega$ the average spectrum vanishes as the chosen grid density.

It is important to realize that this behavior does not depend on our choice of including the width factor from \eqref{Riemann} in the model vector $\bar{f}_n=f_n w_n$ or, $\bar{\mathbf{f}}=\mathbf{Wf}$, where $\mathbf{W}=\mathrm{diag}(\mathbf{w})$. If we include it, instead, in the kernel, the kernel matrix is modified $\bar{\mathbf{K}}:=\mathbf{KW}$, so that $\chi^2(\mathbf{f})=\|\mathbf{g}-\bar{\mathbf{K}}\mathbf{f}\|^2 = \|\mathbf{g}-\mathbf{K}\bar{\mathbf{f}}\|^2=\bar{\chi}^2(\bar{\mathbf{f}})$, and, by a change of variables
\begin{align}\label{Itrafo}
 \bar{\mathbf{f}}_\mathrm{ASM} 
 &= c_{\bar{\chi}^2}\prod_{n=1}^N \!\int_0^\infty\!d\bar{f}_n\,\bar{\mathbf{f}}\,e^{-\frac12\bar{\chi}^2(\bar{\mathbf{f}})}\\
 &= c_{\bar{\chi}^2}\prod_{n=1}^N w_n \prod_{n=1}^N\!\int_0^\infty\!df_n\,\mathbf{Wf}\,e^{-\frac12\chi^2(\mathbf{f})}
  = \mathbf{W}\,\mathbf{f}_\mathrm{ASM}\,, \nonumber
\end{align}
where the constants $w_n$ account for the change in normalization of the Gaussian after the change of variables: $c_{\chi^2}=c_{\bar{\chi}^2}\det(\mathbf{W})=c_{\bar{\chi}^2}\prod_n w_n$.

To understand the grid dependence of $\bar{\mathbf{f}}_\mathrm{ASM}$ we can use a similar argument. Let $\bar{f}_n$ and $\tilde{f}_n$ be the models on two different grids, $\rho(x)$ and $\tilde{\rho}(x)$, that cover the same range, e.g.~$x\in(0,\infty)$, and have the same number of grid points $N$. $\bar{f}_n$ is the integral of the model over the interval $I_n$ centered around $x(z_n)$. Following \eqref{Riemann}, we may express it in terms of the $\tilde{\mathbf{f}}$ as a weighted sum of the $\tilde{f}_{\tilde{n}}$,
defining a linear transformation $\bar{\mathbf{f}}=\tilde{\mathbf{W}} \tilde{\mathbf{f}}$. The situation is quite similar to \eqref{Itrafo}, but with a crucial difference: In general $\tilde{\mathbf{W}}$ will not be diagonal, so that the transformation will change the limits of integration from $\bar{f}_n\ge0$ for $\bar{f}_n$ to $\tilde{\mathbf{W}}\tilde{f}_n\ge0$ for the integration over $\tilde{f}_n$ and consequently $\bar{\mathbf{f}}_\mathrm{ASM}\ne\tilde{\mathbf{W}}{\tilde{\mathbf{f}}}_\mathrm{ASM}$. Apparently, choosing different grids implies different definitions of what values of the model are allowed.

This becomes even more evident when we consider what happens when we refine the grid by halving each interval: Instead of the original $N$ values $\bar{f}_n$ on the original grid, we now have twice as many values $\tilde{f}_{\tilde{n}}$ representing the integral of the model over the halved intervals. The two sets are thus related by $\bar{f}_n=\tilde{f}_{2n-1}+\tilde{f}_{2n}$. 
Sampling the $\tilde{f}_{\tilde{n}}\ge0$ we find that the probability of sampling a given value $\bar{f}_n$ is proportional to
\begin{equation}\label{refine}
 \int_0^\infty\!\!\!d\tilde{f}_{2n-1}\!\int_0^\infty\!\!\!d\tilde{f}_{2n}\,\delta(\bar{f}_n{-}\tilde{f}_{2n-1}{-}\tilde{f}_{2n})
 =\!\int_0^{\bar{f}_n}\!\!\!d\tilde{f}_{2n} = \bar{f}_n ,
\end{equation}
i.e., sampling the $\tilde{f}_{\tilde{n}}$ on the fine grid with a flat distribution implies sampling on the coarse grid with a distribution that is biased against small values of $\tilde{f}_n$.
In other words, the naive discretization of the functional integral \eqref{MCint} does not have a proper continuum limit. We, consequently, have to investigate the definition of a functional integral more carefully.

\subsection{Functional integrals}
We have just seen that the naive discretization of the functional integral, used so successfully in Feynman path integrals \cite{Schulman}, does not work for averaging spectra. The problem is that sampling with a flat distribution on different grids gives incompatible results so that the discretized functional integral has no proper continuum limit \cite{priors}. We can, however, enforce such compatibility in \eqref{refine} by introducing (separate) probability distributions for the $\bar{f}_n$ and the $\tilde{f}_{\tilde{n}}$ on the original and the halved intervals
\begin{equation}\label{compatibility}
 \int_0^{\bar{f}_n}\!\!d\tilde{f}_{2n}\,\tilde{p}(\tilde{f}_{2n})\,\tilde{p}(\bar{f}_n-\tilde{f}_{2n})
 = \bar{p}(\bar{f}_n) .
\end{equation}
In principle, the probability distributions on the two subintervals could be chosen independently, $\tilde{p}_{2n-1}$ and $\tilde{p}_{2n}$. To avoid any bias we assume, however, that the distribution only depends on the width but not the position of the interval. Thus $\tilde{p}_{2n-1}=\tilde{p}_{2n}=:\tilde{p}$, since each subinterval is half the width of the original interval.

The compatibility condition \eqref{compatibility} means that the convolution of $\tilde{p}$ with itself equals $\bar{p}$, which in terms of the Laplace transform
\begin{equation}
 \mathcal{L}\{\bar{p}\}(s) = \int_0^\infty dt\,p(t)\,e^{-st} ,
\end{equation}
is equivalent to $(\mathcal{L}\{\tilde{p}\})^2 = \mathcal{L}\{\bar{p}\}$. To find the compatible distribution on the fine grid given the distribution on the original grid, we just have to take the inverse transform of the square root of its Laplace transform:
$\tilde{p}=\mathcal{L}^{-1}\{\sqrt{\mathcal{L}\{p\}}\}$.

We want the distribution on the original grid to resemble a flat distribution. An obvious choice is to simply introduce a cutoff: $\bar{p}_c(\bar{f})=\big(\Theta(\bar{f})-\Theta(\bar{f}{-}\bar{c})\big)/\bar{c}$, where $\Theta(x)$ is the step function that vanishes for $t{<}0$ and is one for $t{>}0$. The square root of its Laplace transform is $\sqrt{1{-}e^{-\bar{c}s}}/\sqrt{\bar{c}s}$. Expanding the numerator for $s{>}0$ in $e^{-\bar{c}s}$ and using that $\mathcal{L}\{\Theta(t{-}a)/\sqrt{t{-}a}\}(s) = e^{-as}\,\Gamma(\frac12)/\sqrt{s}$, where $\Gamma(z)=\int_0^\infty x^{z-1}e^{-x}\,dx$ is the Gamma function, we find 
\begin{equation}
 \mathcal{L}^{-1}\{\sqrt{\mathcal{L}\{p\}}\}(\tilde{f})
 = \frac1{\sqrt{\pi\bar{c}}}\!\left(\!\frac{\Theta(\tilde{f})}{\sqrt{\tilde{f}}} 
   - \frac12\frac{\Theta(\tilde{f}{-}\bar{c})}{\sqrt{\tilde{f}{-}\bar{c}}} 
   - \cdots \!\right)\!,
\end{equation}
which is negative due to the divergences at integer multiples of the cutoff $\bar{c}$. Thus, for flat distributions $\bar{p}_c(\bar{f})$ with cutoff there exist no compatible distributions $\tilde{p}_c(\tilde{f})$ on the halved intervals. They are called indivisible \cite{priors}.

Alternatively, we can start from an exponential $p_e(\bar{f})=\lambda e^{-\lambda\bar{f}}$, which for $\lambda\searrow0$ approaches a flat distribution. Its Laplace transform is $\mathcal{L}\{\bar{p}_e\}(s)=\lambda/(s+\lambda)$. Using $\mathcal{L}\{e^{-at}/\sqrt{\pi t}\}(s)=(s+a)^{-1/2}$ we see that $\tilde{p}_e(\tilde{f})=e^{-\lambda\tilde{f}}/\sqrt{\pi\tilde{f}/\lambda}$. Thus, the exponential distribution is divisible. 
In fact, from $\mathcal{L}\{f(t)\,e^{-\lambda t}\}(s){=}\mathcal{L}\{f(t)\}(s{+}\lambda)$ and
\begin{equation}
 \mathcal{L}\{t^{\tilde{w}-1}\}(s) = s^{-\tilde{w}}\!\!\int_0^\infty \!\!\!x^{\tilde{w}-1}\,e^{-x}\,dx
 = s^{-\tilde{w}}\,\Gamma(\tilde{w})
\end{equation}
it follows that it can be divided into any number, $n$, of intervals of width $\tilde{w}=1/n$, i.e., it is infinitely divisible. Note that $\tilde{w}$ is the width of the subinterval in units of the width of the original interval. The process of subdivision is consistent: halving the small intervals produces a distribution $\tilde{\tilde{p}}=\mathcal{L}^{-1}\{\sqrt{\mathcal{L}\{\tilde{p}\}}\}$, which, by $\mathcal{L}\{\tilde{p}\}=\sqrt{\mathcal{L}\{\bar{p}\}}$, is equal to $\mathcal{L}^{-1}\{\sqrt[4]{\mathcal{L}\{\bar{p}\}}\}$,
so that the continuum limit of the functional integral is well defined.
Of course, we are not restricted to subintervals of equal width. For
\begin{align}\label{gammadist}
 p_{w,\lambda}(f) 
 &=\mathcal{L}^{-1}\!\left\{\!\sqrt[1/w]{\mathcal{L}\{\lambda e^{-\lambda f}\}}\right\}(f)\\
 &= \frac{\lambda^{w}}{\Gamma(w)}f^{w-1}\,e^{-\lambda f}\nonumber
  =\frac{\;\;f^{w-1}\,e^{-\lambda f}}{\int_0^\infty x^{w-1}\,e^{-\lambda x}\,dx} \,,
\end{align}
which is a gamma distribution with shape parameter $w$ and scale $\lambda$, we find the generalized compatibility relation
\begin{equation}\label{genconsistency}
 \int_0^{\bar{f}}\!d\tilde{f}\,p_{\tilde{w},\lambda}(\tilde{f})\,p_{\bar{w}-\tilde{w},\lambda}(\bar{f}{-}\tilde{f})
 = p_{\bar{w},\lambda}(\bar{f}) \,,
\end{equation}
where the scale $\lambda$ remains unchanged, while the shape parameter changes with the width of the interval. 

Using gamma distributions we can now write down a discretization of the functional integral with a well defined continuum limit. For a particular grid of $N$ points and density $\rho(x)$, we start with the naive discretization \eqref{MCint}, i.e., we sample the $\bar{f}_n$ from a flat distribution
\begin{equation}\label{initI}
  \bar{\mathbf{f}}_\mathrm{ASM} 
 = \lim_{\lambda\to0} c_{\bar{\chi}^2}\prod_{n=1}^N\!\int_0^\infty\!d\bar{f}_n\,\frac{p_{1,\lambda}(\bar{f}_n)}{\lambda}\,\bar{\mathbf{f}}\,e^{-\frac12\bar{\chi}^2(\mathbf{f})} ,
\end{equation}
where convergence in the limit $\lambda\to0$ is guaranteed by the Gaussian.
On a different grid of $\tilde{N}$ points with grid density $\tilde{\rho}(x)$ we then have to sample the $\tilde{f}_{\tilde{n}}$ from a gamma distribution, where the shape parameter is the width of the interval of grid $(\tilde{N},\tilde{\rho}(x))$ in units of the width of the corresponding interval on grid $(N,\rho(x))$. Approximating the width of an interval containing $\tilde{x}$ by $1/N\rho(\tilde{x})$ as in \eqref{Riemann}, we obtain
\begin{equation}\label{compat}
 \tilde{\mathbf{f}}_\mathrm{ASM} \sim
 \prod_{\tilde{n}=1}^{\tilde{N}}\!\int_0^\infty\!d\tilde{f}_{\tilde{n}}\,\tilde{f}_{\tilde{n}}^{\frac{N\rho(\tilde{x}_{\tilde{n}})}{\tilde{N}\tilde{\rho}(\tilde{x}_{\tilde{n}})}-1}\,\tilde{\mathbf{f}}\,e^{-\frac12\tilde{\chi}^2(\tilde{\mathbf{f}})} ,
\end{equation}
which for $\tilde{N}\to\infty$ has a well defined continuum limit, i.e., defines a specific functional integration.

\begin{figure}
\includegraphics[width=0.94\columnwidth]{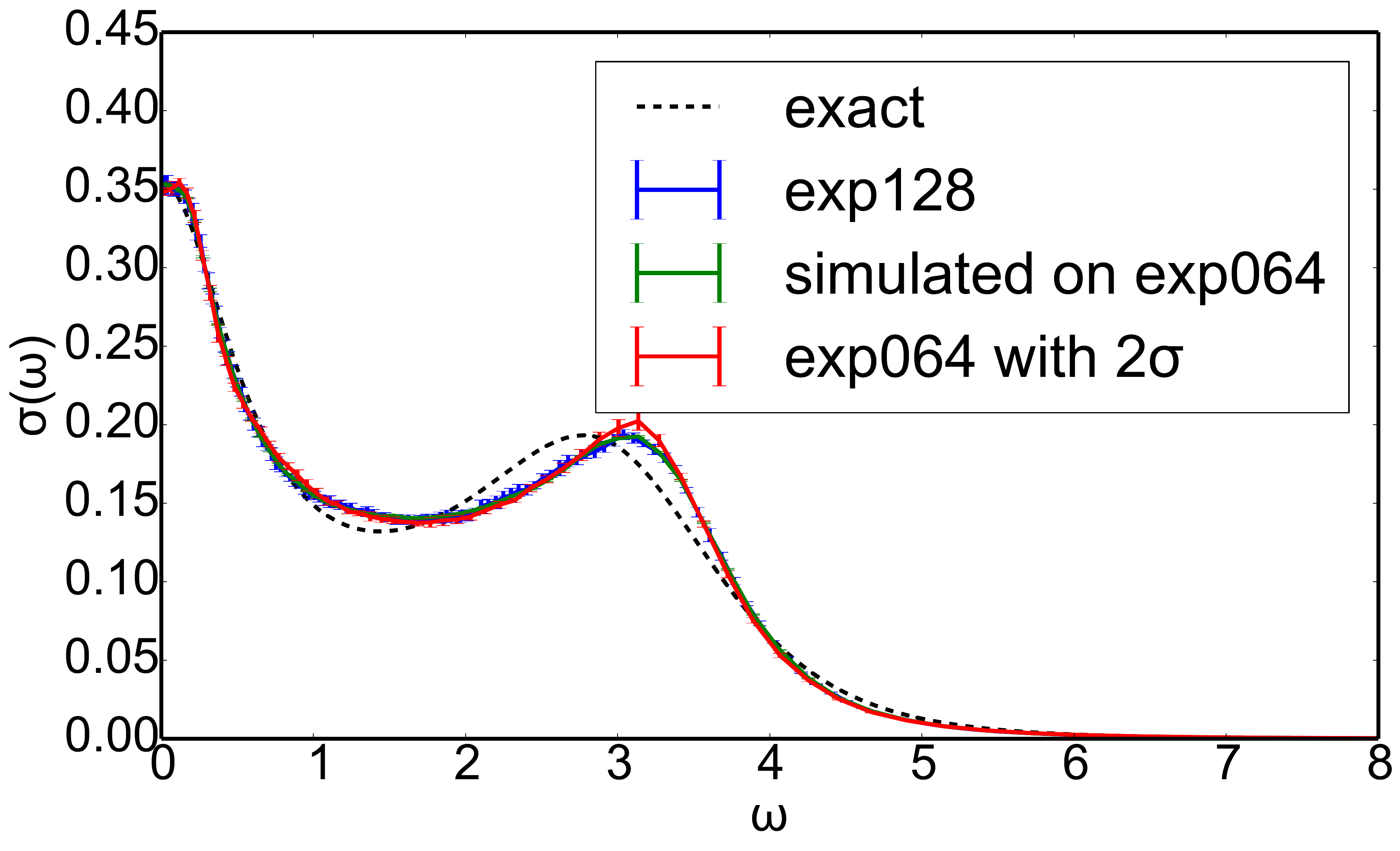}\\[-1ex]\includegraphics[width=0.94\columnwidth]{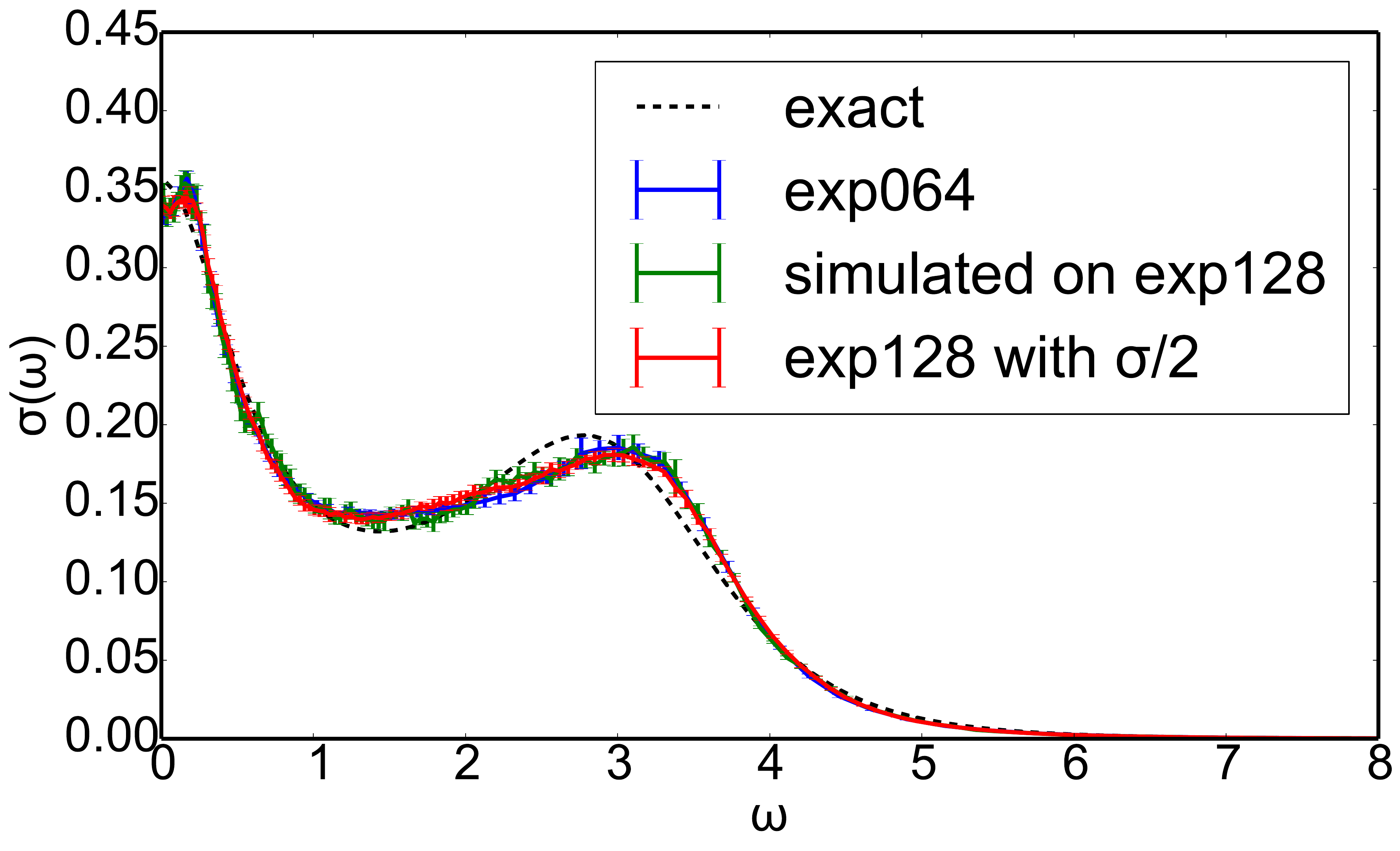}\\[-3ex]
\caption{\label{simulating}(Color online) Simulating one finite grid on another. For the same problem as in the upper panel of Fig.~\ref{grid_dependence} we compare the result for an exponential grid with $\beta{=}3$ and $N{=}128$ points with the simulation on the same exponential but with 
$\tilde{N}{=}64$ (top panel) and vice versa (lower panel). The results agree 
within error bars. 
In addition, runs on the $\tilde{N}$ grid are shown, where the noise in the data is scaled by $N/\tilde{N}$. We do not plot the large symbols distinguishing the curves as they would obscure the near perfect agreement.}
\end{figure}
We can actually use \eqref{compat} to simulate on grid $(\tilde{N},\tilde{\rho}(x))$ the result we would obtain sampling with a flat distribution on a different grid $(N,\rho(x))$. This is illustrated in Fig.~\ref{simulating}.
Note that for $\tilde{N}\tilde{\rho}(\tilde{x}_{\tilde{n}})> N\rho(\tilde{x}_{\tilde{n}})$ the reweighting factor in \eqref{compat} diverges for small $\tilde{f}_{\tilde{n}}$ (but still giving a probability distribution). In the limit $\tilde{N}\to\infty$ individual samples $\tilde{\mathbf{f}}$ will therefore be zero almost everywhere except for finite values on a few intervals, i.e., they will look like a collection of discrete peaks \cite{priors}. This atomicity property of the gamma distributions makes sampling coarse grids on finer ones somewhat noisy.

\begin{figure}
 \includegraphics[width=0.94\columnwidth]{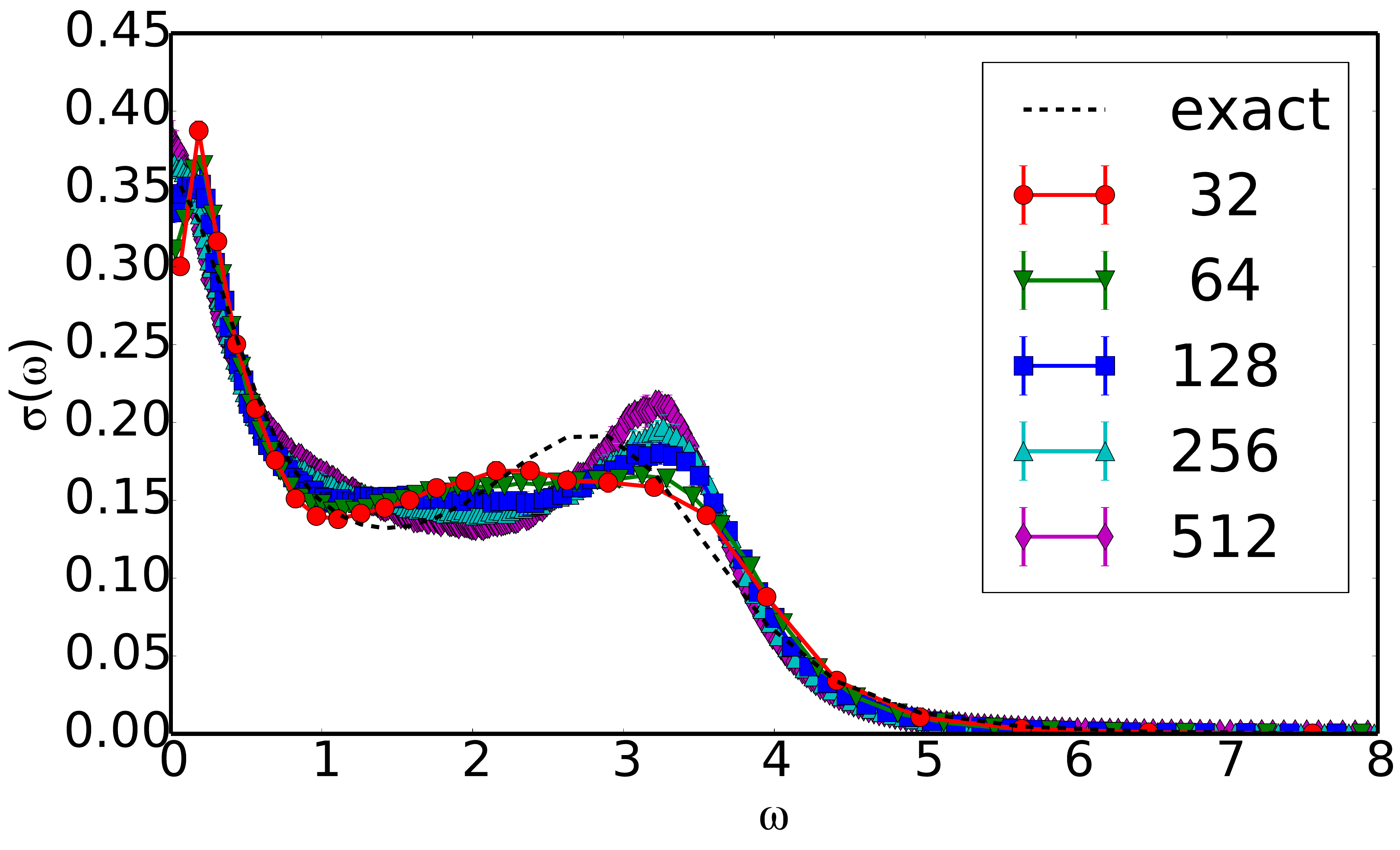}\\[-1ex]
 \includegraphics[width=0.92\columnwidth]{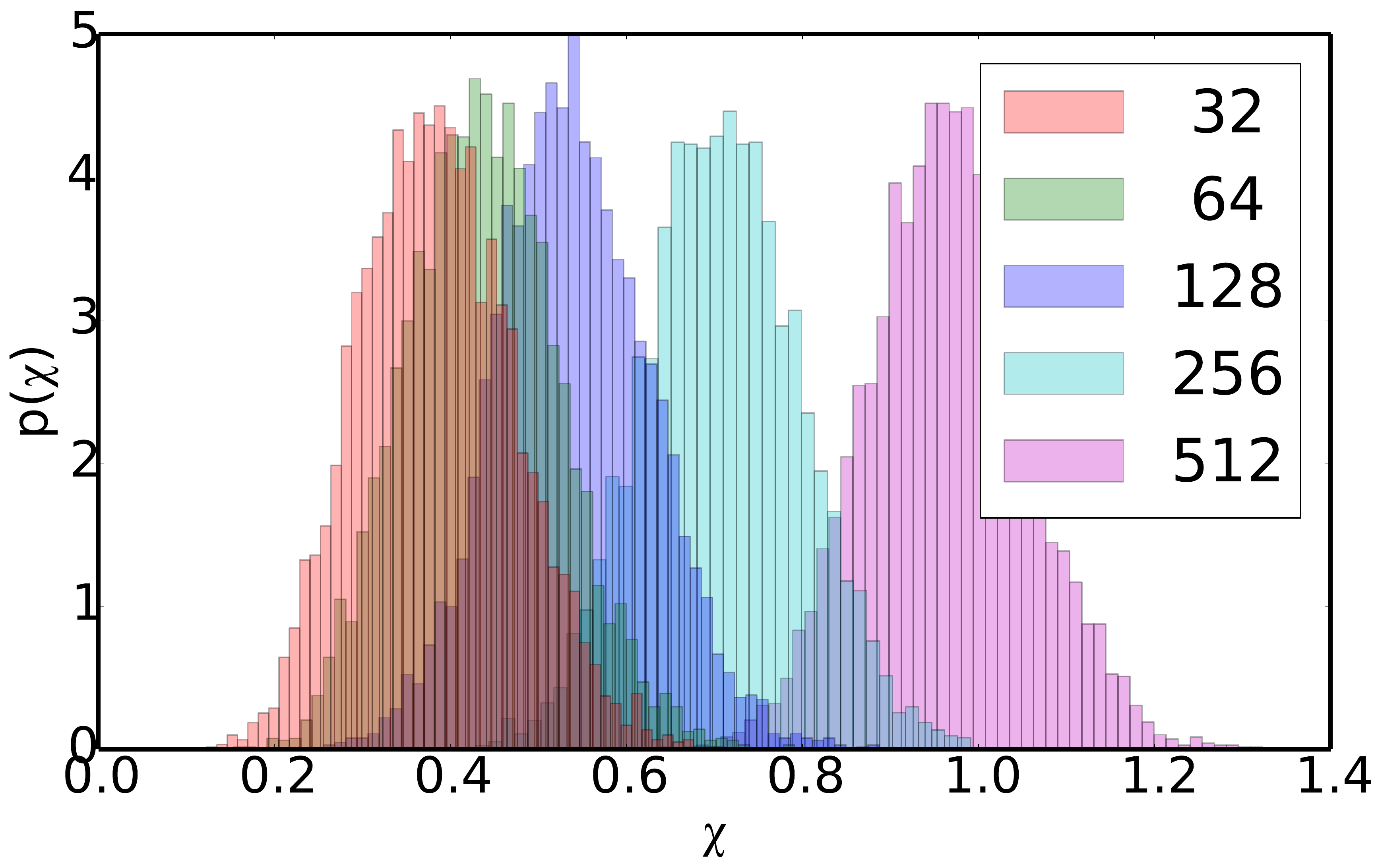}\\[-3ex]
 \caption{\label{NdepBad}(Color online) Dependence of the average spectrum on the number $N$ of grid points for the same problem as in Fig.~\ref{simulating} on a Lorentzian with $\gamma=2.5$. For larger $N$ the average spectra get worse. The reason becomes clear from the histograms in the bottom panel, showing the contribution of spectra with a given fit $\chi$ to $\tilde{\mathbf{f}}_{\mathrm{ASM}}$: with increasing $N$ the histogram moves to the right, i.e., worse fits.}
\end{figure}
Still, we are left with the problem of how to choose the grid $(N,\rho(x))$ used in \eqref{initI}, which determines the functional measure. Our first impulse might be to choose $N$ as large as possible as to minimize discretization errors. As shown in Fig.~\ref{NdepBad}, however, for larger $N$ the average spectra tend to develop spurious structures.
To understand the origin of this counterintuitive behavior, we analyze what models actually contribute to the average $\mathbf{f}_{\mathrm{ASM}}$. Fig.~\ref{NdepBad} shows that with increasing $N$, the average spectrum is eventually dominated by models that fit the data less and less well.
We can understand this qualitatively by realizing that $N$ is the number of degrees of freedom in a model. So increasing $N$ allows for a larger variety of different models. Still, for any given $N$ there is only a single  model that gives the best fit $\chi_{\mathrm{NNLS}}$. Thus the density of models with worse fit increases with $N$, explaining the drift of the histogram towards larger $\chi$.

We can make a more rigorous argument and gain further insights by using the reweighting approach. Let us assume that we are calculating \eqref{initI} on a very fine grid $(N,\rho(x))$. We can simulate the result on a much coarser grid of the same density $(\tilde{N},\rho(x))$ with $\tilde{N}\ll N$. 
Imposing the sum rule $\sum_{\tilde{n}}\tilde{f}_{\tilde{n}}=\tilde{F}$, \eqref{compat} becomes
\begin{equation*}
 \tilde{\mathbf{f}}_\mathrm{ASM}
 \sim\prod_{\tilde{n}=1}^{\tilde{N}}\!\int_0^\infty\!d\tilde{f}_{\tilde{n}}\,\tilde{f}_{\tilde{n}}^{\tilde{w}_{\tilde{n}}-1}
 \,\delta\Big(\tilde{F}-\sum_{\tilde{n}}\tilde{f}_{\tilde{n}}\Big)
 \,\tilde{\mathbf{f}}\,e^{-\frac12\tilde{\chi}^2(\tilde{\mathbf{f}})} 
\end{equation*}
with $\tilde{w}_{\tilde{n}}{:=}N/\tilde{N}\gg1$. The models are thus sampled from a Dirichlet distribution  
\begin{equation}\label{simD}
 p_D(\tilde{w}_1,\ldots,\tilde{w}_{\tilde{N}}; \tilde{f}_1,\ldots,\tilde{f}_{\tilde{N}}) \!= \frac{\Gamma\big(\sum\tilde{w}_{\tilde{n}}\big)}{\tilde{F}^{\sum\tilde{w}_{\tilde{n}}{-}1}\!\prod\!\Gamma(\tilde{w}_{\tilde{n}})} \prod\tilde{f}_{\tilde{n}}^{\tilde{w}_{\tilde{n}}{-}1}
\end{equation}
with fixed $\tilde{F}$, 
where the normalization constant
\begin{equation*}
\int\limits_0^{\tilde{F}} d\tilde{f}_1\tilde{f}_1^{\tilde{w}_1-1} \hspace{-2.5ex}
      \int\limits_0^{\tilde{F}-\tilde{f}_1}\hspace{-1.5ex} d\tilde{f}_2\tilde{f}_2^{\tilde{w}_2-1} \hspace{-1ex} \ldots \hspace{-4.5ex} 
      \int\limits_0^{\tilde{F}-\sum_{\tilde{n}=1}^{\tilde{N}-2}\tilde{f}_{\tilde{n}}}\hspace{-4.5ex} d\tilde{f}_{\tilde{N}-1}\tilde{f}_{\tilde{N}{-}1}^{\tilde{w}_{\tilde{N}\!{-}1}\!{-}1}
  \Big(\!\sum_{\tilde{n}=1}^{\tilde{N}-1}\!\tilde{f}_{\tilde{n}}-\tilde{F}\!\Big)^{\tilde{w}_{\tilde{N}}}
\end{equation*}
follows from Euler's Beta integral \cite{DLMF} for $\alpha, \beta>0$
\begin{equation}
 \int_0^1 t^{\alpha-1}(1-t)^{\beta-1}\,dt = \frac{\Gamma(\alpha)\Gamma(\beta)}{\Gamma(\alpha+\beta)}\,.
\end{equation}
For $\tilde{w}_{\tilde{n}}\gg1$, using Stirling's formula $\ln\Gamma(z)\approx z\ln z-z$ 
we obtain
\begin{equation}
 \ln p_D \approx
 \sum\!\left(\!\tilde{w}_{\tilde{n}}\ln\frac{\sum_{\tilde{n}}\tilde{w}_{\tilde{n}}}{\tilde{w}_{\tilde{n}}}\frac{\tilde{f}_{\tilde{n}}}{\tilde{F}}\!\right)\!
 =-\frac{N}{\tilde{F}}\sum_{\tilde{n}}\tilde{f}\ln\frac{\tilde{f}}{\tilde{f}_{\tilde{n}}}
\end{equation}
which is proportional to the entropy $S(\tilde{f}|\tilde{\mathbf{f}})$ of 
$\tilde{f}:=\tilde{F}/\tilde{N}$ relative to $\tilde{f}_{\tilde{n}}$. 
Hence
\begin{equation}\label{AvgEnt}
 \tilde{\mathbf{f}}_{\mathrm{ASM}}\sim 
 \prod_{\tilde{n}=1}^{\tilde{N}}\!\int_0^\infty\!\!\!\!d\tilde{f}_{\tilde{n}}
 \,\delta\Big(\tilde{F}{-}\!\sum_{\tilde{n}}\tilde{f}_{\tilde{n}}\Big)
 \,\tilde{\mathbf{f}}\,e^{\frac{N}{\tilde{F}}S(\tilde{f}|\,\tilde{\mathbf{f}})-\frac12\tilde{\chi}^2(\tilde{\mathbf{f}})} .
\end{equation}
For $N\to\infty$ the entropy term will dominate $\chi^2$ so that the integrals of the model over the intervals, $\tilde{\mathbf{f}}_{\mathrm{ASM}}$, will tend to a constant, independent of the data. The situation is quite analogous to that discussed for \eqref{def}: Sampling on a very dense grid gives a model proportional to the grid density, which, again, acts as a default model.

In fact, the prior on the models in \eqref{AvgEnt} is strikingly similar to the maximum entropy prior, which, however, uses the entropy of the model relative to the default model. The two relative entropies are closely related, with the MaxEnt entropy $-\sum\tilde{f}_{\tilde{n}}\ln\tilde{f}_{\tilde{n}}/\tilde{f}$ penalizing models deviating from the default somewhat less than the average-spectrum entropy $-\sum\tilde{f}\ln\tilde{f}/\tilde{f}_{\tilde{n}}$. 

While the grid density acts as a default model, the number $N$ of grid points plays the role of a regularization parameter: going from $N$ to $N'$ grid points changes the prefactor of the entropy term relative to that of the fit function by $N'/N$. In \eqref{AvgEnt} we can reach the same effect by staying with the $N$ grid points but scaling the fit function by $N/N'$, i.e., scaling the overall variance in the data. Fig.~\ref{simulating} shows that this is a simple, efficient, and remarkably accurate way of simulating grids with the same density but different number of points. 
This explains why the idea of rescaling the noise of Monte Carlo data is widely used in practice \cite{Sandvik98,Gunnarsson07,Fuchs10,Sandvik16}.
Moreover, it beautifully confirms the intuition underlying the idea of the average spectrum method stated after eqn.~\eqref{PI}: the noise in the data leads, via the averaging of spectra, to a smoothing of the model, and the larger the noise, the larger this regularizing effect.

\section{Practical method}

To make the average spectrum approach a practical method, we have to understand how to choose the regularization. As we have seen in Fig.~\ref{NdepBad}, results can depend strongly on the number of grid points. Most striking about this dependence is that with increasing regularization the average spectra are not smoothed but rather develop increasingly sharp features---the opposite of what one would expect from a regularization!
To understand this, we look at how the default model fits the imaginary-axis data. For this we need to relate the grid density to the default model. We can, e.g., write the default optical conductivity as $\sigma_\text{def}(\omega):=\pi\Pi(0)\rho(\omega)/2$ from which we can calculate the values on the imaginary axis as
\begin{equation}
 \Pi_\text{def}(\omega_m) = \Pi(0)\left(1-\int_0^\infty\frac{\rho(\omega)}{1+(\omega/\omega_m)^2}\,d\omega\right) .
\end{equation}
\begin{figure}[t]
 \includegraphics[width=0.94\columnwidth]{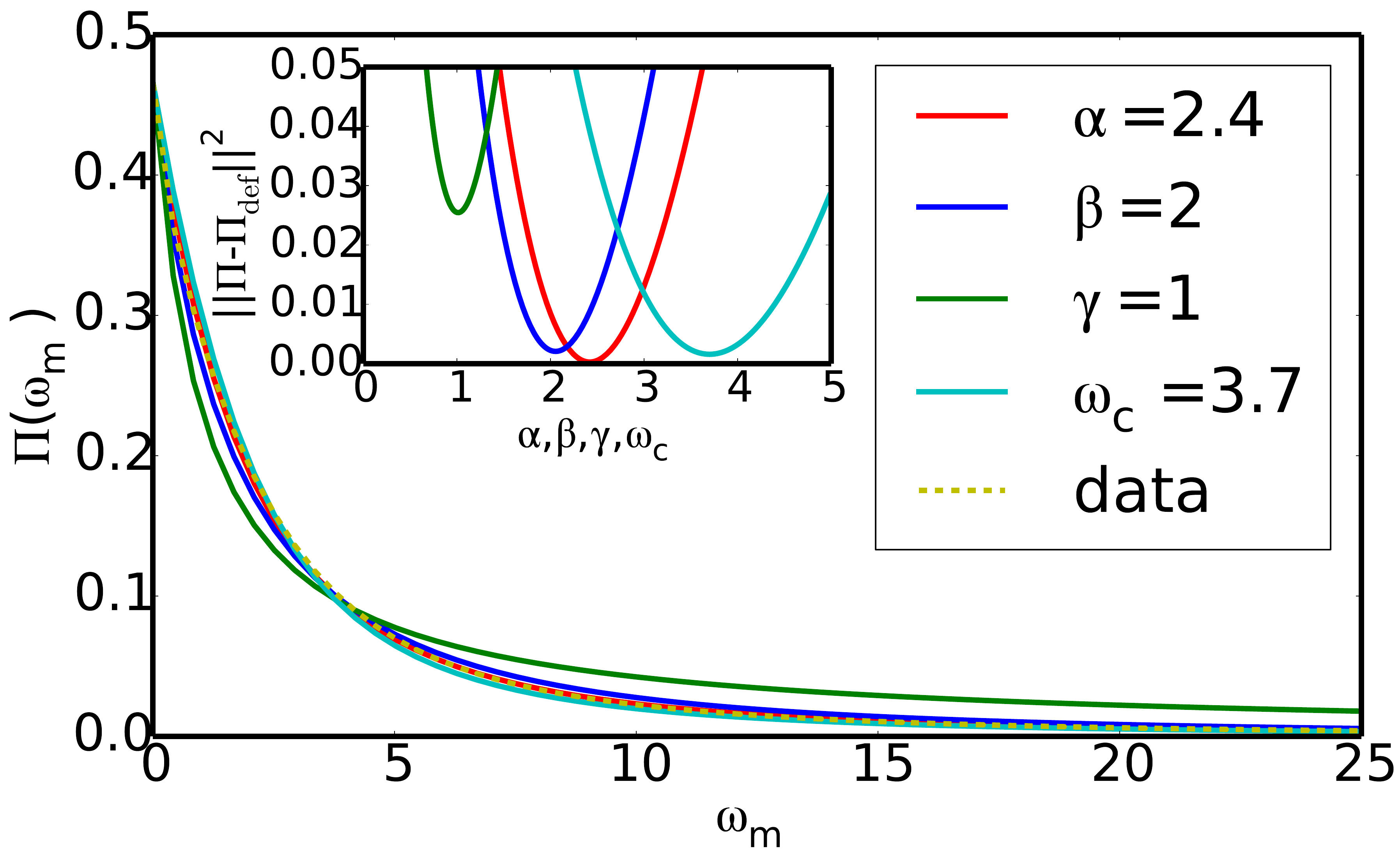}
 \caption{\label{fitgrid}(Color Online) Determining a reasonable default model by fitting the grid density $\rho(x)$ (Gaussian of standard deviation $\alpha$, exponential with rate $\beta$, or Lorentzian of width $\gamma$, uniform with cutoff $\omega_c$) to the data (dotted line) for the same problem as in Fig.~\ref{NdepBad}. The inset shows $\sum \big(\Pi(\omega_m)-\Pi_\text{def}(\omega_m)\big)^2$ as a function of the grid parameter, highlighting the importance of choosing a reasonable default model.}
\end{figure}

The deviation of $\Pi_\text{def}$ from the actual data tells us how compatible the default model is with the data. This is shown in Fig.~\ref{fitgrid}. We find that the grid density used in Fig.~\ref{NdepBad}, a Lorentzian of width $\gamma=2.5$, does not even remotely represent the imaginary-axis data. The situation is even worse for the uniform grids of Fig.~\ref{cutoff_dependence}, which, with increasing cutoff, become more and more inconsistent with the data. In all these cases the default model does not resemble the data on the imaginary axis at all. This misfit has dramatic consequences, since the information we try to extract from the data is hidden in the tiny details on the imaginary axis---the very reason why analytic continuation is so ill-conditioned. Regularizing towards a grossly wrong default model then forces the models to develop unphysical features in order to somehow achieve a decent fit nevertheless. 

\begin{figure}
 \includegraphics[width=0.94\columnwidth]{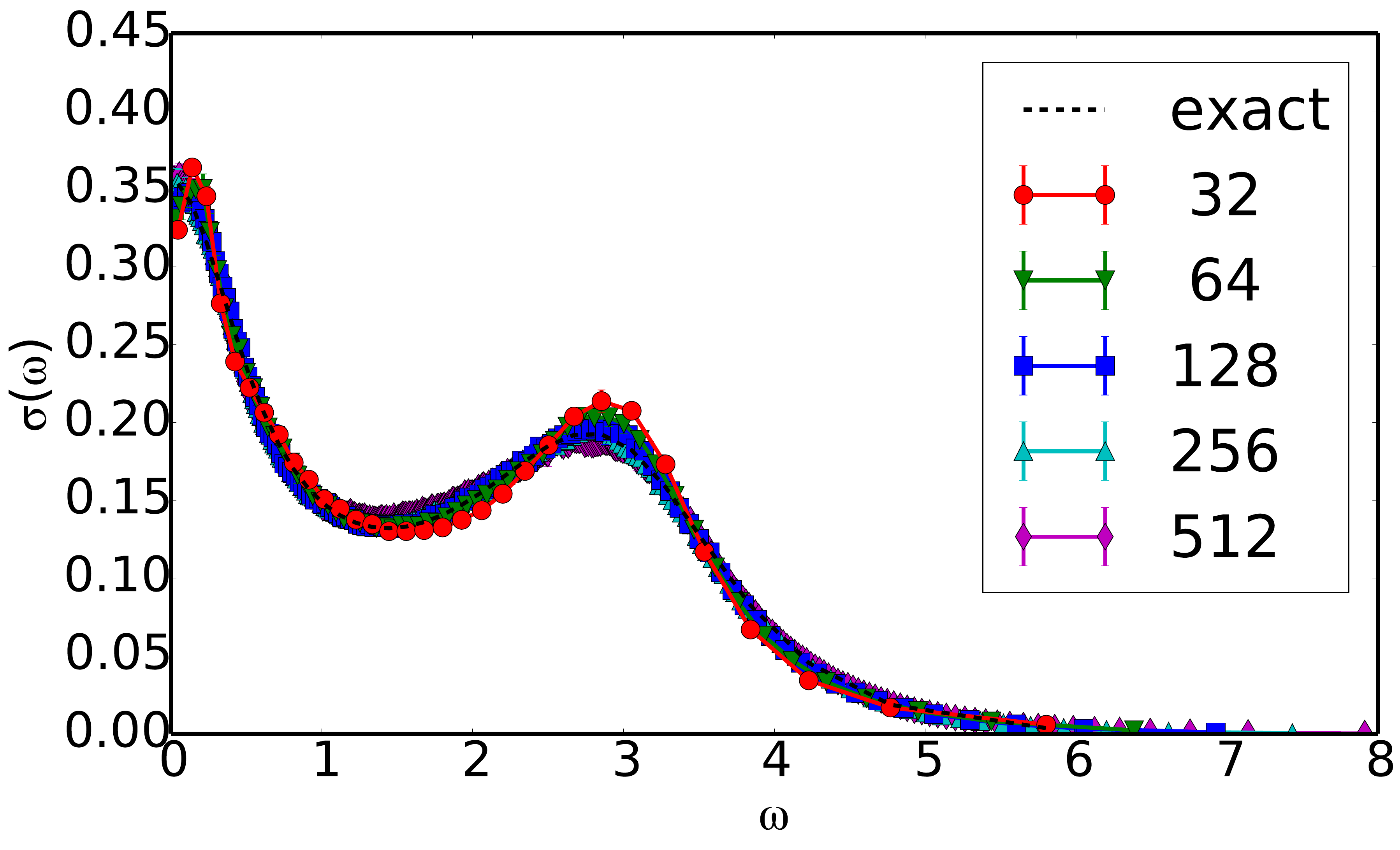}\\[-1ex]
 \includegraphics[width=0.92\columnwidth]{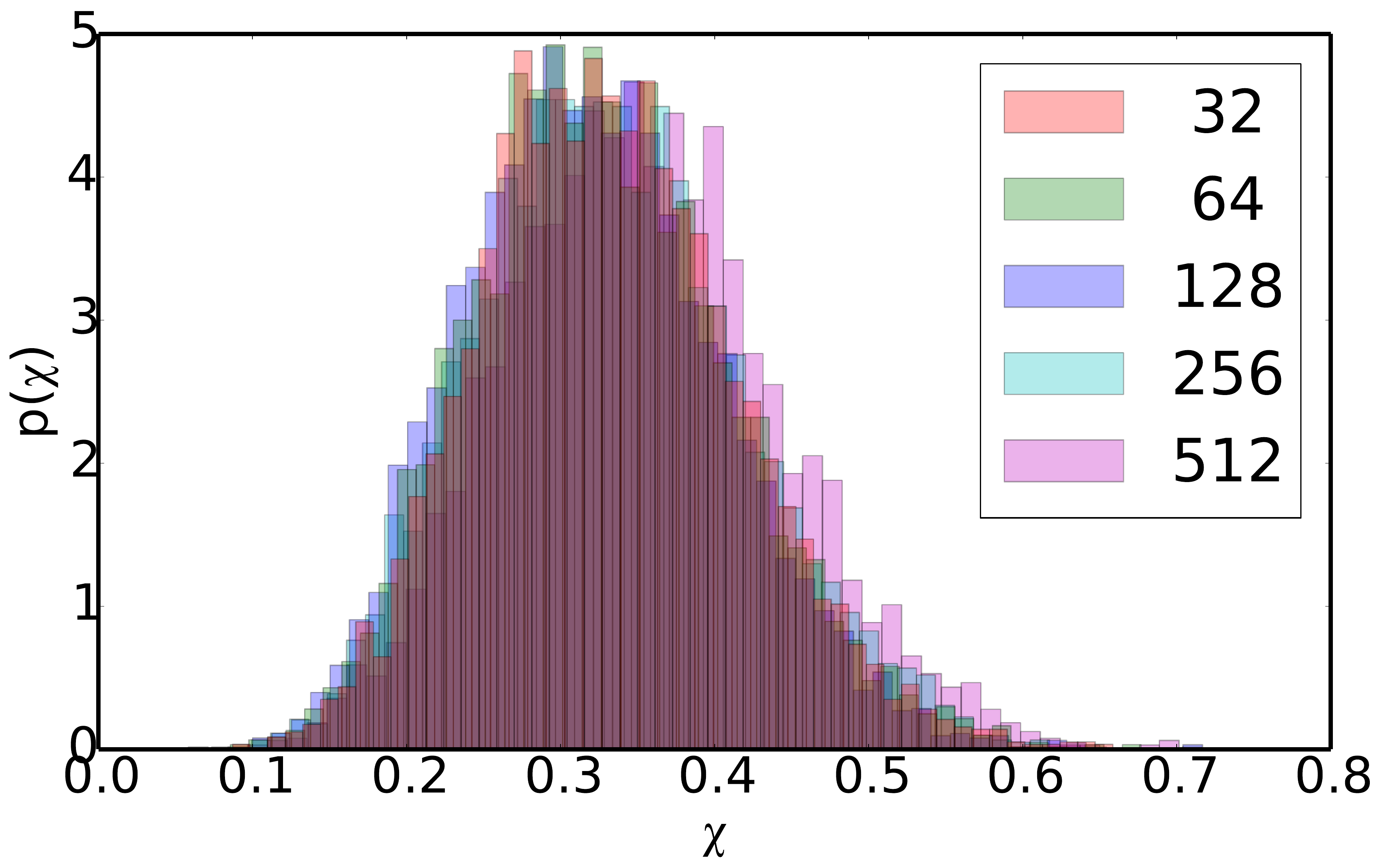}\\[-3ex]
 \caption{\label{NdepGood}(Color online) Dependence of the average spectrum on the number $N$ of grid points for the same problem as in Fig.~\ref{NdepBad} on an optimized Gaussian grid with $\alpha=2.4$. The average spectra are largely independent of $N$ and the histograms show a consistently good fit.}
\end{figure}

The problem completely disappears when using a reasonable default model. An example is shown in Fig.~\ref{NdepGood}, using a Gaussian grid with $\alpha=2.4$. As we read off from Fig.~\ref{fitgrid}, this default model is compatible with the data and we see that with increasing regularization the resulting spectra become smoother. Moreover, this smoothing is not very strong so that the results are remarkably robust under changes in the number of grid points.
It thus turns out that the choice of the default model is much more important than that of the regularization parameter.

In this respect, a flat default model with cutoff is a particularly unfortunate choice. As we see from Fig.~\ref{fitgrid}, for a cutoff $\omega_c\approx3.7$ we actually obtain quite a reasonable default model so that we would expect robust average spectra. Such a grid, however, has no points in the tail of the model. If we want to resolve the model at higher frequencies we need to ``improve'' the cutoff, necessarily giving increasingly poor default models that are responsible for the disastrous results obtained in Fig.~\ref{cutoff_dependence}.

\section{Conclusions}

We have seen that the average spectrum method is not the parameter free method suggested by the deceptively written functional integral \eqref{PI}: We have to choose a grid density $\rho(x)$, which acts as a default model, and a number $N$ of grid points, which acts as a regularization parameter. The reason for this is that the naive discretization \eqref{MCint} does not converge to a well defined functional integral. Instead we have to sample the components of the models we are integrating over from distributions that are consistent for different discretizations. For general non-negative functions these are gamma distributions \eqref{gammadist}, when, in addition, the functions fulfill a sum-rule they are Dirichlet distributions \eqref{simD}. 
This raises, of course, the question why the naive discretization does work for path integrals. In the Feynman approach the integrand itself already fulfills the consistency relation giving rise to a complex Wiener measure \cite{Gelfand60}, so that the appropriate functional measure is inherent in the path integrand. 
This is not the case for the functional integral \eqref{PI}, requiring us to explicitly specify the functional measure by singling out a specific grid on which to evaluate \eqref{MCint}. Using the corresponding family of gamma or Dirichlet distributions, 
we can then take the continuum limit.

We find that approaching this limit we sample models $\tilde{\mathbf{f}}$ with a prior given by the entropy of the flat distribution on that grid relative to $\tilde{\mathbf{f}}$, making the grid density act as a default model, while the number of grid points acts as the regularization parameter.
The similarity with the maximum-entropy method (MaxEnt) is obvious. Of course, the entropies differ, but this only means that MaxEnt regularizes large deviations from the default model somewhat less. More importantly, MaxEnt determines the model from maximizing rather than averaging. This appears to avoid having to specify a functional measure. But, in fact, in the derivation of MaxEnt marginalizations over the model space do require functional integrals. To quote \cite{Sivia}, p.~137: ``This shortcoming has been missed earlier due to a deceptive side-effect of the Gaussian approximation made in the calculation, and because the quantitative answers from the analysis were generally sensible in practice.''

To make the average spectrum method a practical technique for analytic continuation we need reliable recipes for choosing grid density and number of grid points. As we have demonstrated, the results can depend quite strongly on these choices. 
A badly chosen default model will bias the results towards models that give an extremely bad fit to the imaginary-axis data. In such cases we obtain utterly unreasonable results: with increasing regularization the result develops stronger and stronger features.
Interestingly this is particularly true for flat default models with cutoff, which are by their very nature ill suited for analytic continuation.
A good default model should, instead, not only be featureless but also be overall consistent with the data. For such default models the features in the results will be suppressed with increasing regularization---as it should be. In fact, then results become fairly independent of the actual choice of the regularization parameter over a wide range, highlighting the importance of the default model rather than the regularization parameter.

Finally, a practical method must be efficient. This has so far been the cardinal problem of the average spectrum method. We have described an optimized implementation, without which we could not have analyzed the method in such detail. While we have discussed here only one specific test case, more can be found in \cite{GhanemPhD}. In addition we make an efficient web-based implementation freely available at \url{www.spektra.app}.


\begin{thebibliography}{25}%
\makeatletter
\providecommand \@ifxundefined [1]{%
 \@ifx{#1\undefined}
}%
\providecommand \@ifnum [1]{%
 \ifnum #1\expandafter \@firstoftwo
 \else \expandafter \@secondoftwo
 \fi
}%
\providecommand \@ifx [1]{%
 \ifx #1\expandafter \@firstoftwo
 \else \expandafter \@secondoftwo
 \fi
}%
\providecommand \natexlab [1]{#1}%
\providecommand \enquote  [1]{``#1''}%
\providecommand \bibnamefont  [1]{#1}%
\providecommand \bibfnamefont [1]{#1}%
\providecommand \citenamefont [1]{#1}%
\providecommand \href@noop [0]{\@secondoftwo}%
\providecommand \href [0]{\begingroup \@sanitize@url \@href}%
\providecommand \@href[1]{\@@startlink{#1}\@@href}%
\providecommand \@@href[1]{\endgroup#1\@@endlink}%
\providecommand \@sanitize@url [0]{\catcode `\\12\catcode `\$12\catcode
  `\&12\catcode `\#12\catcode `\^12\catcode `\_12\catcode `\%12\relax}%
\providecommand \@@startlink[1]{}%
\providecommand \@@endlink[0]{}%
\providecommand \url  [0]{\begingroup\@sanitize@url \@url }%
\providecommand \@url [1]{\endgroup\@href {#1}{\urlprefix }}%
\providecommand \urlprefix  [0]{URL }%
\providecommand \Eprint [0]{\href }%
\providecommand \doibase [0]{http://dx.doi.org/}%
\providecommand \selectlanguage [0]{\@gobble}%
\providecommand \bibinfo  [0]{\@secondoftwo}%
\providecommand \bibfield  [0]{\@secondoftwo}%
\providecommand \translation [1]{[#1]}%
\providecommand \BibitemOpen [0]{}%
\providecommand \bibitemStop [0]{}%
\providecommand \bibitemNoStop [0]{.\EOS\space}%
\providecommand \EOS [0]{\spacefactor3000\relax}%
\providecommand \BibitemShut  [1]{\csname bibitem#1\endcsname}%
\let\auto@bib@innerbib\@empty
%</preamble>
\bibitem [{\citenamefont {Jarrell}\ and\ \citenamefont
  {Gubernatis}(1996)}]{MaxEntReview96}%
  \BibitemOpen
  \bibfield  {author} {\bibinfo {author} {\bibfnamefont {M.}~\bibnamefont
  {Jarrell}}\ and\ \bibinfo {author} {\bibfnamefont {J.~E.}\ \bibnamefont
  {Gubernatis}},\ }\href {\doibase 10.1016/0370-1573(95)00074-7} {\bibfield
  {journal} {\bibinfo  {journal} {Phys. Rep.}\ }\textbf {\bibinfo {volume}
  {269}},\ \bibinfo {pages} {133} (\bibinfo {year} {1996})}\BibitemShut
  {NoStop}%
\bibitem [{\citenamefont {Reymbaut}\ \emph {et~al.}(2015)\citenamefont
  {Reymbaut}, \citenamefont {Bergeron},\ and\ \citenamefont
  {Tremblay}}]{Tremblay15}%
  \BibitemOpen
  \bibfield  {author} {\bibinfo {author} {\bibfnamefont {A.}~\bibnamefont
  {Reymbaut}}, \bibinfo {author} {\bibfnamefont {D.}~\bibnamefont {Bergeron}},
  \ and\ \bibinfo {author} {\bibfnamefont {A.-M.~S.}\ \bibnamefont
  {Tremblay}},\ }\href {\doibase 10.1103/PhysRevB.92.060509} {\bibfield
  {journal} {\bibinfo  {journal} {Phys. Rev. B}\ }\textbf {\bibinfo {volume}
  {92}},\ \bibinfo {pages} {060509(R)} (\bibinfo {year} {2015})}\BibitemShut
  {NoStop}%
\bibitem [{\citenamefont {Burnier}\ and\ \citenamefont
  {Rothkopf}(2013)}]{QCD13}%
  \BibitemOpen
  \bibfield  {author} {\bibinfo {author} {\bibfnamefont {Y.}~\bibnamefont
  {Burnier}}\ and\ \bibinfo {author} {\bibfnamefont {A.}~\bibnamefont
  {Rothkopf}},\ }\href {\doibase 10.1103/PhysRevLett.111.182003} {\bibfield
  {journal} {\bibinfo  {journal} {Phys. Rev. Lett.}\ }\textbf {\bibinfo
  {volume} {111}},\ \bibinfo {pages} {182003} (\bibinfo {year}
  {2013})}\BibitemShut {NoStop}%
\bibitem [{\citenamefont {Hansen}(2010)}]{Hansen}%
  \BibitemOpen
  \bibfield  {author} {\bibinfo {author} {\bibfnamefont {P.~C.}\ \bibnamefont
  {Hansen}},\ }\href {\doibase 10.1137/1.9780898718836} {\emph {\bibinfo
  {title} {Discrete Inverse Problems}}}\ (\bibinfo  {publisher} {SIAM},\
  \bibinfo {year} {2010})\BibitemShut {NoStop}%
\bibitem [{\citenamefont {Jarrell}(2012)}]{JarrellCorrel12}%
  \BibitemOpen
  \bibfield  {author} {\bibinfo {author} {\bibfnamefont {M.}~\bibnamefont
  {Jarrell}},\ }in\ \href {http://www.cond-mat.de/events/correl12/manuscripts}
  {\emph {\bibinfo {booktitle} {Correlated Electrons: From Models to
  Materials}}},\ \bibinfo {editor} {edited by\ \bibinfo {editor} {\bibfnamefont
  {E.}~\bibnamefont {Pavarini}}, \bibinfo {editor} {\bibfnamefont
  {E.}~\bibnamefont {Koch}}, \bibinfo {editor} {\bibfnamefont {F.}~\bibnamefont
  {Anders}}, \ and\ \bibinfo {editor} {\bibfnamefont {M.}~\bibnamefont
  {Jarrell}}}\ (\bibinfo  {publisher} {Forschungszentrum J\"ulich},\ \bibinfo
  {year} {2012})\BibitemShut {NoStop}%
\bibitem [{\citenamefont {White}(1991)}]{White91}%
  \BibitemOpen
  \bibfield  {author} {\bibinfo {author} {\bibfnamefont {S.~R.}\ \bibnamefont
  {White}},\ }in\ \href {\doibase 10.1007/978-3-642-76382-3_13} {\emph
  {\bibinfo {booktitle} {Computer Simulation Studies in Condensed Matter
  Physics {III}}}},\ \bibinfo {editor} {edited by\ \bibinfo {editor}
  {\bibfnamefont {D.~P.}\ \bibnamefont {Landau}}, \bibinfo {editor}
  {\bibfnamefont {K.~K.}\ \bibnamefont {Mon}}, \ and\ \bibinfo {editor}
  {\bibfnamefont {B.-B.}\ \bibnamefont {Sch\"uttler}}}\ (\bibinfo  {publisher}
  {Springer},\ \bibinfo {address} {Heidelberg},\ \bibinfo {year} {1991})\ pp.\
  \bibinfo {pages} {145--153}\BibitemShut {NoStop}%
\bibitem [{\citenamefont {Sandvik}(1998)}]{Sandvik98}%
  \BibitemOpen
  \bibfield  {author} {\bibinfo {author} {\bibfnamefont {A.~W.}\ \bibnamefont
  {Sandvik}},\ }\href {\doibase 10.1103/PhysRevB.57.10287} {\bibfield
  {journal} {\bibinfo  {journal} {Phys. Rev. B}\ }\textbf {\bibinfo {volume}
  {57}},\ \bibinfo {pages} {10287} (\bibinfo {year} {1998})}\BibitemShut
  {NoStop}%
\bibitem [{\citenamefont {Vafayi}\ and\ \citenamefont
  {Gunnarsson}(2007)}]{Gunnarsson07}%
  \BibitemOpen
  \bibfield  {author} {\bibinfo {author} {\bibfnamefont {K.}~\bibnamefont
  {Vafayi}}\ and\ \bibinfo {author} {\bibfnamefont {O.}~\bibnamefont
  {Gunnarsson}},\ }\href {\doibase 10.1103/PhysRevB.76.035115} {\bibfield
  {journal} {\bibinfo  {journal} {Phys. Rev. B}\ }\textbf {\bibinfo {volume}
  {76}},\ \bibinfo {pages} {035115} (\bibinfo {year} {2007})}\BibitemShut
  {NoStop}%
\bibitem [{\citenamefont {Sylju\r{a}sen}(2008)}]{Syljuasen08}%
  \BibitemOpen
  \bibfield  {author} {\bibinfo {author} {\bibfnamefont {O.~F.}\ \bibnamefont
  {Sylju\r{a}sen}},\ }\href {\doibase 10.1103/physRevB.78.174429} {\bibfield
  {journal} {\bibinfo  {journal} {Phys. Rev. B}\ }\textbf {\bibinfo {volume}
  {78}},\ \bibinfo {pages} {174429} (\bibinfo {year} {2008})}\BibitemShut
  {NoStop}%
\bibitem [{\citenamefont {Fuchs}\ \emph {et~al.}(2010)\citenamefont {Fuchs},
  \citenamefont {Pruschke},\ and\ \citenamefont {Jarrell}}]{Fuchs10}%
  \BibitemOpen
  \bibfield  {author} {\bibinfo {author} {\bibfnamefont {S.}~\bibnamefont
  {Fuchs}}, \bibinfo {author} {\bibfnamefont {T.}~\bibnamefont {Pruschke}}, \
  and\ \bibinfo {author} {\bibfnamefont {M.}~\bibnamefont {Jarrell}},\ }\href
  {\doibase 10.1103/PhysRevE.81.056701} {\bibfield  {journal} {\bibinfo
  {journal} {Phys. Rev. E}\ }\textbf {\bibinfo {volume} {81}},\ \bibinfo
  {pages} {056701} (\bibinfo {year} {2010})}\BibitemShut {NoStop}%
\bibitem [{\citenamefont {Beach}(2004)}]{Beach04}%
  \BibitemOpen
  \bibfield  {author} {\bibinfo {author} {\bibfnamefont {K.~S.~D.}\
  \bibnamefont {Beach}},\ }\href {https://arxiv.org/abs/cond-mat/0403055}
  {\enquote {\bibinfo {title} {Identifying the maximum entropy method as a
  special limit of stochastic analytic continuation},}\ } (\bibinfo {year}
  {2004}),\ \Eprint {http://arxiv.org/abs/cond-mat/0403055} {cond-mat/0403055}
  \BibitemShut {NoStop}%
\bibitem [{\citenamefont {Sandvik}(2016)}]{Sandvik16}%
  \BibitemOpen
  \bibfield  {author} {\bibinfo {author} {\bibfnamefont {A.~W.}\ \bibnamefont
  {Sandvik}},\ }\href {\doibase 10.1103/PhysRevE.94.063308} {\bibfield
  {journal} {\bibinfo  {journal} {Phys. Rev. E}\ }\textbf {\bibinfo {volume}
  {94}},\ \bibinfo {pages} {063308} (\bibinfo {year} {2016})}\BibitemShut
  {NoStop}%
\bibitem [{\citenamefont {Boehnke}\ \emph {et~al.}(2011)\citenamefont
  {Boehnke}, \citenamefont {Hafermann}, \citenamefont {Ferrero}, \citenamefont
  {Lechermann},\ and\ \citenamefont {Parcollet}}]{Legendre}%
  \BibitemOpen
  \bibfield  {author} {\bibinfo {author} {\bibfnamefont {L.}~\bibnamefont
  {Boehnke}}, \bibinfo {author} {\bibfnamefont {H.}~\bibnamefont {Hafermann}},
  \bibinfo {author} {\bibfnamefont {M.}~\bibnamefont {Ferrero}}, \bibinfo
  {author} {\bibfnamefont {F.}~\bibnamefont {Lechermann}}, \ and\ \bibinfo
  {author} {\bibfnamefont {O.}~\bibnamefont {Parcollet}},\ }\href {\doibase
  10.1103/PhysRevB.84.075145} {\bibfield  {journal} {\bibinfo  {journal} {Phys.
  Rev. B}\ }\textbf {\bibinfo {volume} {84}},\ \bibinfo {pages} {075145}
  (\bibinfo {year} {2011})}\BibitemShut {NoStop}%
\bibitem [{{\relax DLMF}()}]{DLMF}%
  \BibitemOpen
  {\relax DLMF},\ \href {http://dlmf.nist.gov/} {\enquote {\bibinfo {title}
  {{\it NIST Digital Library of Mathematical Functions}},}\ }\bibinfo
  {howpublished} {http://dlmf.nist.gov/, Release 1.0.17 of 2017-12-22},\
  \bibinfo {note} {{F.~W.~J. Olver, A.~B. Olde Daalhuis, D.~W. Lozier, B.~I.
  Schneider, R.~F. Boisvert, C.~W. Clark, B.~R. Miller and B.~V. Saunders,
  eds.}}\BibitemShut {Stop}%
\bibitem [{\citenamefont {Gunnarsson}\ \emph {et~al.}(2010)\citenamefont
  {Gunnarsson}, \citenamefont {Haverkort},\ and\ \citenamefont
  {Sangiovanni}}]{Gunnarsson10}%
  \BibitemOpen
  \bibfield  {author} {\bibinfo {author} {\bibfnamefont {O.}~\bibnamefont
  {Gunnarsson}}, \bibinfo {author} {\bibfnamefont {M.~W.}\ \bibnamefont
  {Haverkort}}, \ and\ \bibinfo {author} {\bibfnamefont {G.}~\bibnamefont
  {Sangiovanni}},\ }\href {\doibase 10.1103/PhysRevB.82.165125} {\bibfield
  {journal} {\bibinfo  {journal} {Phys. Rev. B}\ }\textbf {\bibinfo {volume}
  {82}},\ \bibinfo {pages} {165125} (\bibinfo {year} {2010})}\BibitemShut
  {NoStop}%
\bibitem [{\citenamefont {Robert}(1995)}]{TruncNormal95}%
  \BibitemOpen
  \bibfield  {author} {\bibinfo {author} {\bibfnamefont {C.~P.}\ \bibnamefont
  {Robert}},\ }\href {\doibase 10.1007/BF00143942} {\bibfield  {journal}
  {\bibinfo  {journal} {Statistics and Computing}\ }\textbf {\bibinfo {volume}
  {5}},\ \bibinfo {pages} {121} (\bibinfo {year} {1995})}\BibitemShut {NoStop}%
\bibitem [{\citenamefont {Lawson}\ and\ \citenamefont
  {Hanson}(1974)}]{LawsonHanson}%
  \BibitemOpen
  \bibfield  {author} {\bibinfo {author} {\bibfnamefont {C.~L.}\ \bibnamefont
  {Lawson}}\ and\ \bibinfo {author} {\bibfnamefont {R.~J.}\ \bibnamefont
  {Hanson}},\ }\href {\doibase 10.1137/1.9781611971217} {\emph {\bibinfo
  {title} {Solving Least Squares Problems}}}\ (\bibinfo  {publisher} {SIAM},\
  \bibinfo {year} {1974})\BibitemShut {NoStop}%
\bibitem [{\citenamefont {Krivenko}\ and\ \citenamefont
  {Rubtsov}(2006)}]{Rubtsov06}%
  \BibitemOpen
  \bibfield  {author} {\bibinfo {author} {\bibfnamefont {I.~S.}\ \bibnamefont
  {Krivenko}}\ and\ \bibinfo {author} {\bibfnamefont {A.~N.}\ \bibnamefont
  {Rubtsov}},\ }\href {https://arxiv.org/abs/cond-mat/0612233} {\enquote
  {\bibinfo {title} {Analytic continuation of quantum {Monte Carlo} data:
  Optimal stochastic regularization approach},}\ } (\bibinfo {year} {2006}),\
  \Eprint {http://arxiv.org/abs/cond-mat/0612233} {cond-mat/0612233}
  \BibitemShut {NoStop}%
\bibitem [{\citenamefont {Arsenault}\ \emph {et~al.}(2017)\citenamefont
  {Arsenault}, \citenamefont {Neuberg}, \citenamefont {Hannah},\ and\
  \citenamefont {Millis}}]{Millis16}%
  \BibitemOpen
  \bibfield  {author} {\bibinfo {author} {\bibfnamefont {L.-F.}\ \bibnamefont
  {Arsenault}}, \bibinfo {author} {\bibfnamefont {R.}~\bibnamefont {Neuberg}},
  \bibinfo {author} {\bibfnamefont {L.~A.}\ \bibnamefont {Hannah}}, \ and\
  \bibinfo {author} {\bibfnamefont {A.~J.}\ \bibnamefont {Millis}},\ }\href
  {\doibase https://doi.org/10.1088/1361-6420/aa8d93} {\bibfield  {journal}
  {\bibinfo  {journal} {Inverse Problems}\ }\textbf {\bibinfo {volume} {33}},\
  \bibinfo {pages} {115007} (\bibinfo {year} {2017})}\BibitemShut {NoStop}%
\bibitem [{\citenamefont {Waldvogel}(2010)}]{Waldvogel}%
  \BibitemOpen
  \bibfield  {author} {\bibinfo {author} {\bibfnamefont {J.}~\bibnamefont
  {Waldvogel}},\ }in\ \href@noop {} {\emph {\bibinfo {booktitle} {Approximation
  and Computation}}},\ \bibinfo {series} {Springer Optimization and Its
  Applications}, Vol.~\bibinfo {volume} {42},\ \bibinfo {editor} {edited by\
  \bibinfo {editor} {\bibfnamefont {W.}~\bibnamefont {Gautschi}}, \bibinfo
  {editor} {\bibfnamefont {G.}~\bibnamefont {Mastroianni}}, \ and\ \bibinfo
  {editor} {\bibfnamefont {T.}~\bibnamefont {Rassias}}}\ (\bibinfo  {publisher}
  {Springer},\ \bibinfo {address} {New York, NY},\ \bibinfo {year} {2010})\
  pp.\ \bibinfo {pages} {267--282}\BibitemShut {NoStop}%
\bibitem [{\citenamefont {Schulman}(2005)}]{Schulman}%
  \BibitemOpen
  \bibfield  {author} {\bibinfo {author} {\bibfnamefont {L.~S.}\ \bibnamefont
  {Schulman}},\ }\href@noop {} {\emph {\bibinfo {title} {Techniques and
  Applications of Path Integration}}}\ (\bibinfo  {publisher} {Dover
  Publications},\ \bibinfo {year} {2005})\BibitemShut {NoStop}%
\bibitem [{\citenamefont {Skilling}\ and\ \citenamefont
  {Sibisi}(1996)}]{priors}%
  \BibitemOpen
  \bibfield  {author} {\bibinfo {author} {\bibfnamefont {J.}~\bibnamefont
  {Skilling}}\ and\ \bibinfo {author} {\bibfnamefont {S.}~\bibnamefont
  {Sibisi}},\ }\enquote {\bibinfo {title} {Priors on measures},}\ in\
  \href@noop {} {\emph {\bibinfo {booktitle} {Maximum Entropy and {Bayesian}
  Methods}}},\ \bibinfo {editor} {edited by\ \bibinfo {editor} {\bibfnamefont
  {K.~M.}\ \bibnamefont {Hanson}}\ and\ \bibinfo {editor} {\bibfnamefont
  {R.~N.}\ \bibnamefont {Silver}}}\ (\bibinfo  {publisher} {Kluwer},\ \bibinfo
  {year} {1996})\ pp.\ \bibinfo {pages} {261--270}\BibitemShut {NoStop}%
\bibitem [{\citenamefont {Gel'fand}\ and\ \citenamefont
  {Yaglom}(1960)}]{Gelfand60}%
  \BibitemOpen
  \bibfield  {author} {\bibinfo {author} {\bibfnamefont {I.~M.}\ \bibnamefont
  {Gel'fand}}\ and\ \bibinfo {author} {\bibfnamefont {A.~M.}\ \bibnamefont
  {Yaglom}},\ }\href {\doibase 10.1063/1.1703636} {\bibfield  {journal}
  {\bibinfo  {journal} {J. Math. Phys.}\ }\textbf {\bibinfo {volume} {1}},\
  \bibinfo {pages} {48} (\bibinfo {year} {1960})}\BibitemShut {NoStop}%
\bibitem [{\citenamefont {Sivia}\ and\ \citenamefont {Skilling}(2006)}]{Sivia}%
  \BibitemOpen
  \bibfield  {author} {\bibinfo {author} {\bibfnamefont {C.~S.}\ \bibnamefont
  {Sivia}}\ and\ \bibinfo {author} {\bibfnamefont {J.}~\bibnamefont
  {Skilling}},\ }\href@noop {} {\emph {\bibinfo {title} {Data Analysis: A
  Bayesian Tutorial}}},\ \bibinfo {edition} {2nd}\ ed.\ (\bibinfo  {publisher}
  {Oxford University Press},\ \bibinfo {year} {2006})\BibitemShut {NoStop}%
\bibitem [{\citenamefont {Ghanem}(2017)}]{GhanemPhD}%
  \BibitemOpen
  \bibfield  {author} {\bibinfo {author} {\bibfnamefont {K.}~\bibnamefont
  {Ghanem}},\ }\emph {\bibinfo {title} {Stochastic Analytic Continuation: A
  Bayesian Approach}},\ \href@noop {} {Ph.D. thesis},\ \bibinfo  {school} {RWTH
  Aachen University} (\bibinfo {year} {2017})\BibitemShut {NoStop}%
\end{thebibliography}
\end{document}